\documentclass[12pt]{article}

\begin{document}


\newcommand{\bb}{\begin{equation}}
\newcommand{\ee}{\end{equation}}
\newcommand{\bbb}{\begin{eqnarray}}
\newcommand{\eee}{\end{eqnarray}}
\newcommand{\vc}[1]{\mbox{$\vec{{\bf #1}}$}}
\newcommand{\mc}[1]{\mathcal{#1}}
\newcommand{\del}{\partial}
\newcommand{\lk}{\left}
\newcommand{\ave}[1]{\mbox{$\langle{#1}\rangle$}}
\newcommand{\re}{\right}
\newcommand{\pd}[1]{\frac{\del}{\del #1}}
\newcommand{\pdd}[2]{\frac{\del^2}{\del #1 \del #2}}
\newcommand{\Dd}[1]{\frac{d}{d #1}}
\newcommand{\pref}[1]{(\ref{#1})}

\newcommand
{\sect}[1]{\vspace{20pt}{\LARGE}\noindent
{\bf #1:}}
\newcommand
{\subsect}[1]{\vspace{10pt}{\Large}\noindent
{\bf #1:}}

\def\ie{{\it i.e.}}
\def\eg{{\it e.g.}}
\def\cf{{\it c.f.}}
\def\etal{{\it et.al.}}
\def\etc{{\it etc.}}

\def\AA{{\cal A}}
\def\BB{{\cal B}}
\def\CC{{\cal C}}
\def\DD{{\cal D}}
\def\EE{{\cal E}}
\def\FF{{\cal F}}
\def\GG{{\cal G}}
\def\HH{{\cal H}}
\def\II{{\cal I}}
\def\JJ{{\cal J}}
\def\KK{{\cal K}}
\def\LL{{\cal L}}
\def\MM{{\cal M}}
\def\NN{{\cal N}}
\def\OO{{\cal O}}
\def\PP{{\cal P}}
\def\QQ{{\cal Q}}
\def\RR{{\cal R}}
\def\SS{{\cal S}}
\def\TT{{\cal T}}
\def\UU{{\cal U}}
\def\VV{{\cal V}}
\def\WW{{\cal W}}
\def\XX{{\cal X}}
\def\YY{{\cal Y}}
\def\ZZ{{\cal Z}}

\def\inbar{\,\vrule height1.5ex width.4pt depth0pt}
\def\IR{\relax{\rm I\kern-.18em R}}
\def\IC{\relax\hbox{$\inbar\kern-.3em{\rm C}$}}
\def\IH{\relax{\rm I\kern-.18em H}}
\def\IP{\relax{\rm I\kern-.18em P}}
\def\Z{{\bf Z}}
\def\One{{1\hskip -3pt {\rm l}}}
\def\nth{$n^{\rm th}$}

\def\sinh{{\rm sinh}}
\def\cosh{{\rm cosh}}
\def\tanh{{\rm tanh}}
\def\sgn{{\rm sgn}}
\def\det{{\rm det}}
\def\exp{{\rm exp}}
\def\sh{{\rm sh}}
\def\ch{{\rm ch}}

\def\ell{{\it l}}
\def\str{{\it str}}
\def\lp{\ell_{{\rm pl}}}
\def\ls{\ell_{{\str}}}
\def\gs{g_\str}
\def\gym{g_{Y}}
\def\geff{g_{\rm eff}}
\def\eff{{\rm eff}}
\def\r11{R_{11}}
\def\alp{{\alpha'}}
\def\tgs{{\tilde{g}_s}}
\def\talp{{{\tilde{\alpha}}'}}
\def\tlp{{\tilde{\ell}_{{\rm pl}}}}
\def\tr11{{\tilde{R}_{11}}}
\def\wtilde{\widetilde}
\def\what{\widehat}
\def\hlp{{\hat{\ell}_{{\rm pl}}}}
\def\hr11{{\hat{R}_{11}}}
\def\hf{{\textstyle\frac12}}
\def\coeff#1#2{{\textstyle{#1\over#2}}}

\def\CY{Calabi-Yau}

\def\lessapprox{\;{\buildrel{<}\over{\scriptstyle\sim}}\;}
\def\greaterapprox{\;{\buildrel{>}\over{\scriptstyle\sim}}\;}

\input{epsf}

\begin{titlepage}
\rightline{EFI-99-2}

\rightline{hep-th/9901135}

\vskip 3cm
\centerline{\Large{\bf Black Holes and Five-brane Thermodynamics}}

\vskip 2cm
\centerline{
Emil Martinec\footnote{\texttt{ejm@theory.uchicago.edu}}~~ and ~~
Vatche Sahakian\footnote{\texttt{isaak@theory.uchicago.edu}}}
\vskip 12pt
\centerline{\sl Enrico Fermi Inst. and Dept. of Physics}
\centerline{\sl University of Chicago}
\centerline{\sl 5640 S. Ellis Ave., Chicago, IL 60637, USA}

\vskip 2cm

\begin{abstract}
The phase diagram for $Dp$-branes in M-theory
compactified on $T^4$, $T^4/Z_2$, $T^5$, and $T^6$ is constructed.
As for the lower-dimensional tori considered in 
our previous work (hep-th/9810224), the black brane phase
at high entropy connects onto matrix theory at low entropy;
we thus recover all known instances of matrix theory
as consequences of the Maldacena conjecture.
The difficulties that arise for $T^6$ are reviewed.
We also analyze the $D1$-$D5$ system on $T^5$; we exhibit its relation
to matrix models of $M5$-branes, and use spectral flow 
as a tool to investigate the dependence of the
phase structure on angular momentum.  
\end{abstract}

\end{titlepage}

\newpage
\setcounter{page}{1}

\section{Summary of results and discussion}

\subsection{Introductory remarks}

Black hole thermodynamics has played an important role in
elucidating the structure of M-theory 
(see \cite{GARYREVIEW,JUANREVIEW,PEETREVIEW} for reviews).
In the context of the Maldacena conjecture~\cite{MALDA1,WITHOLO,GKP},
black hole thermodynamics generates predictions for the
thermodynamics of gauge theory in various strong-coupling regimes.  
This conjecture posits (in its extended form) that all of M-theory
in spacetimes with particular asymptotic boundary conditions
is equivalent (dual) to a theory without gravity.
Recently \cite{MSSYM123}, the present authors constructed a phase diagram for
maximally supersymmetric Yang-Mills theory (SYM)
on tori $T^p$, $p=1,2,3$, by systematically exploiting these ideas.
It was seen that a number of different geometrical phases
(\ie\ those with a valid low-energy supergravity description
as black objects)
arise as the entropy and coupling are varied
\footnote{The entropy is most useful in parameterizing
the behavior of the theory since it
is directly tied to the horizon area of the low-energy supergravity
solution.  The energy can then be read from the equation
of state of the relevant black hole.}.
The boundaries of the region of geometrical phases are
correspondence curves~\cite{HORPOLCHCORR} , 
where the curvature of the geometry
becomes string scale at the horizon of the object.

Generically, the thermodynamics at high entropy contains a phase
of black Dp-branes, while at low entropy one finds
eleven-dimensional black holes in the light-cone (LC) frame.
The reason is quite simple \cite{MSSYM123}: 
{\it The scaling limit specified by Maldacena,
and the limit prescribed by Sen and Seiberg for compactifications
of matrix theory~\cite{WHYSEIB,ASHOKE}, are one and the same}
\footnote{It was shown that these two limits are related
in \cite{HYUN,HYUNKIEM}.  Demonstrating their complete 
equivalence requires further specifying the dimensionless
quantities to be held fixed, in particular the scale of
the energy.}.
One and the same gauge theory describe both; for example,
on $T^p$, black Dp-branes characterize the density of
states in the regime of SYM entropies $S\greaterapprox N^2$,
whereas matrix theory~\cite{MAT1,MAT2} describes the regime $S<N$.

The point is that the scales of various features
of the geometry, for instance proper size of the torus
and the string coupling, depend on radial position in
the associated low-energy black supergravity solution.  Since 
the horizon radius decreases with decreasing entropy, and
only the horizon geometry is relevant to the thermodynamics,
the entropy parameterizes a path through the moduli space
of the low-energy supergravity.  Along this path, it may
be necessary to perform U-duality transformations to
achieve a valid low-energy description of the horizon geometry.
This is why, at high entropy, the charge carried by the system
is Dp-brane number; while at low entropy, it is interpreted as momentum.  
The two lie on an orbit of the U-duality group $E_p(\Z)$.
Furthermore, phase transitions may occur in the
geometrical region due to (de)localization
of the horizon on cycles across which it is initially (un)smeared
~\cite{LAFLAMME2}.
Such transitions are involved in the passage from
black Dp-branes to Matrix theory black holes 
\cite{BFKS1,BFKS3,HORMART,BFKS2}.

Thermodynamics is one of many probes of Matrix/Maldacena duality.
It is a particularly useful one in that it canonically associates
an energy scale (that of a typical Hawking quantum)
with a particular place in the geometry (the horizon).
The fact that the geometry appropriate to the description of
this scale undergoes a sequence of duality transformations 
as we go from IR (Matrix theory regime) to UV (Maldacena regime) means that
the interpretation of probes as scattering states in DLCQ M-theory
is only valid up to some scale, beyond which one should pass
to a description in terms of scattering off of black $p$-branes
in a dual geometry.  Using the relation between the energy
and the radial scale probed \cite{PEETPOLCH},
this implies that matrix theory is only valid (in the sense
of accurately describing flat-space supergravity) up to some distance
from the source
\footnote{This conclusion was independently reached from a somewhat
different perspective in \cite{SUSSKINDFLAT}.  The analysis of
supersymmetric quantum mechanics (SQM) in
this latter work is equivalent to the large $V$ limit of the phase
diagrams here and in \cite{MSSYM123}.  In section 2.4 of \cite{MSSYM123}
it was observed that the D0 geometry breaks down at
the correspondence point, where the temperature 
of the system is $T\sim N^{1/3} \r11/\lp^2$.
Using the energy-distance relations of \cite{PEETPOLCH},
the result $r_{\rm max}\sim N^{1/3}\lp$ follows.}.

The precise relation between the Maldacena or near-horizon limit
of $N$ Dp-branes on $T^p$, and Matrix theory on $T^p$ 
with $N$ units of longitudinal momentum, goes as follows \cite{MSSYM123}:
The Maldacena limit is 
$\alpha'=\ls^2\rightarrow 0$,
with the gauge coupling $\gym^2=\gs\ls^{(p-3)/2}$ and the 
{\it coordinate} size $\Sigma_i$ of the torus cycles held fixed.
This limit isolates the gauge theory dynamics on the Dp-brane
while decoupling gravity (for $p<6$).
Natural energy scales in the gauge theory are measured
with respect to the torus size
\footnote{For $p\ne3$, the Yang-Mills coupling is dimensionful,
and should be referred to the torus scale as well.
When we say that a dimensionful quantity is held fixed in
the decoupling limit, we mean the energy in
the system relative to that scale.}.
On the other hand, the Seiberg-Sen prescription 
for matrix theory on $T^p$~\cite{WHYSEIB,ASHOKE}
involves IIA string theory with $N$ D0-branes, or
equivalently M-theory with $N$ units of momentum on a circle
of radius $R_{11}$; then one takes the limit
$\lp\rightarrow 0$, with $R_{11}/\lp^2$ and the (transverse)
torus cycle sizes $R_i/\lp$ held fixed.
The relation between the two sets of parameters is simply 
(\cf~\cite{HYUN,HYUNKIEM,MSSYM123})
the T-duality on all cycles of $T^p$ that maps Dp-branes 
to D0 branes and vice-versa:
\bbb
  \ls^2& = &\frac{\lp^3}{R_{11}} \nonumber\\
  \Sigma_i& = &\frac{\lp^3}{\r11 R_i} \nonumber\\
  \gs& = &\left(\frac{\lp}{\r11}\right)^{\frac{p-3}2}
	\prod_{i=1}^p\frac{\lp}{R_i} \nonumber\\
  \gym^2& = &\left(\frac{\lp^2}{\r11}\right)^{p-3}
	\prod_{i=1}^p\frac{\lp}{R_i}\ .
\label{matrixmalda}
\eee
Thus, the two limits are clearly identical.

In this work, we extend our analysis of such compactifications
to $p=4,5$, where the relevant theories involve the dynamics
of five-branes~\cite{BRS,SEIBLITTLE,WHYSEIB,ASHOKE}; 
and $p=6$, where the definition of matrix
theory is problematic~\cite{WHYSEIB,ASHOKE,%
BRUNNERKARCH,HANANYLIFSCHYTZ,LOSMOORES}.
In the process of generating
the phase diagram, we will rediscover all the remaining prescriptions
for generating matrix theory compactifications; we will also
comment on the difficulties encountered for $p=6$ (and a 
proposal by Kachru \etal\ \cite{KLS}
for overcoming them).  For $p=5$, we map out the phase diagram
of the six-dimensional `little string theories' compactified
on a five-torus $T^5$.

In addition, we will analyze the phase diagram of the D1-D5 system,
which arises in diverse contexts:
\begin{itemize}

\item 
It has played a central role in our understanding of black
hole thermodynamics~\cite{STROMVAFA}; 

\item
It is a prime example of the Maldacena conjecture,
due to the rich algebraic structure of 
1+1d superconformal theories which are proposed duals to string theory on 
$AdS_3\times S^3\times \MM_4$~\cite{MALDA1,MALDASTROMADS3,MARTINECADSMM,GKS}; 

\item
It describes the `little string' theory of fivebranes
~\cite{SEIBLITTLE,DVVM5}, 
where the little strings carry both winding and momentum charges.

\item
It is related to the DLCQ description of fivebrane 
dynamics~\cite{ABKSS,SETHISEIBERG,SETHI,GANORSETHI}.
\end{itemize}

\noindent
The analysis will clarify the relation of the D-brane description
of the system to one in terms of NS fivebranes and fundamental
strings~\cite{GKS}, as low-energy descriptions of
different regions of the phase diagram
(for earlier work, see \cite{JOHNSON}).
Finally, we will explore the use of spectral flow
in the superconformal theory to determine the spectral density
of the theory as a function of angular momentum on $S^3$.

\subsection{Phase diagrams for $T^4$, $T^5$, and $T^6$}

As in \cite{MSSYM123}, the phase diagrams for Dp branes on tori, $p=4,5,6$,
have a number of common features.  
The vertical axis of the diagrams will be entropy; for the
horizontal axis we take the size $V$ of cycles on the torus $T^p$ 
in eleven dimensional Planck
units, as measured in the LC M-theory appearing in the
lower right corner (the phase of boosted 11d black holes).
$N$ is the charge carried by the system: brane number in
the high entropy regimes and longitudinal
momentum in the low-energy, LC M-theory phase.
Throughout the various phases, the corresponding gravitational
couplings vanish in the Maldacena limit (except for $p=6$,
where the limit keeps the Planck scale of the high-entropy
phase held fixed), implying the decoupling
of gravity for the dual dynamics.
Solid lines on the diagrams denote thermodynamic transitions separating
distinct phases, while dotted lines represent
symmetry transformations which change the appropriate 
low-energy description. 
We do not expect sharp phase transitions along these dotted
curves since the scaling of 
the equations of state is unchanged across them
\footnote{This, does not in principle exclude 
the possibility of smoother (\ie\ higher order) transitions.}.

\begin{figure}[p]
\epsfxsize=14cm \centerline{\leavevmode \epsfbox{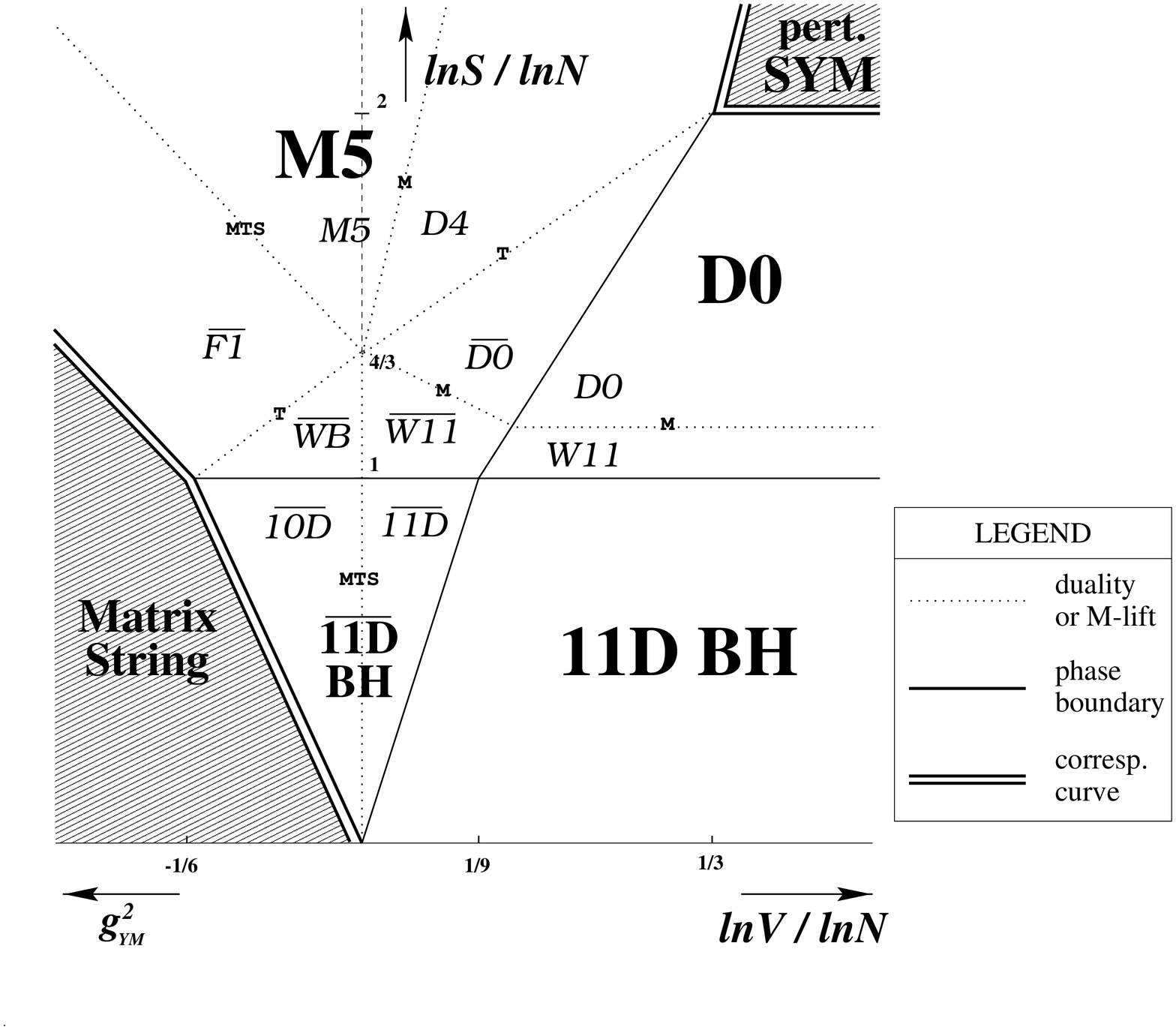}}
\caption{{\small \sl Phase diagram of the six-dimensional
$(2,0)$ theory on $T^4\times S^1$.
$S$ is entropy, $V=R/\lp$ is the size of a cycle on the $T^4$
of light-cone M theory, and $N$ is longitudinal momentum quantum.
The dotted lines denote symmetry transformations:
M: M lift or reduction; T: T duality; S: S duality.
The solid lines are phase transition curves. Double solid lines
denote correspondence curves.
The dashed line is the extension of the axis $V=1$, and
is merely included to help guide the eye.
The label dictionary is as follows:
$D0$: black D0 geometry;
$W11$: black 11D wave geometry;
$11DBH$: 11D LC black hole;
$\overline{D0}$: black smeared D0 geometry;
$\overline{W11}$: black smeared 11D wave geometry;
$\overline{11D}BH$: 11D smeared LC black hole;
$D4$: black D4 geometry;
$M5$: black M5 geometry;
$\overline{F1}$: black smeared fundamental string geometry;
$\overline{WB}$: black smeared IIB wave geometry;
$\overline{10D}BH$: IIB boosted black hole.
The phase diagram can also be considered that of the 
$(2,0)$ theory on $T^4/Z_2\times S^1$
by reinterpreting the 
$\overline{F1}$, $\overline{WB}$, $\overline{10D}$ phases,
and the Matrix string phase as those of a Heterotic theory.
.}}
\label{SYM4fig}
\end{figure}

The structure of the phase diagram for $V>1$ is identical to the cases
encountered in~\cite{MSSYM123} (see, for example, Figure \ref{SYM4fig}). 
At high entropies and large M theory $T^p$, 
we have a perturbative p+1d SYM gas phase.  Its Yang-Mills
coupling $\gym$ increases toward the left, 
\cf\ Equation~\pref{matrixmalda}.
The effective dimensionless coupling is of order one 
on the double lines bounding this phase, 
which are Horowitz-Polchinski correspondence curves.
As the entropy decreases at large $V$,
there is a D0 brane phase arising on the right and middle of the diagrams.
Its description as a thermodynamic state within
SYM theory would be highly interesting. It has
a Horowitz-Polchinski correspondence point at $S\sim N^2$, where a zero
specific heat transition is to occur~\cite{BKR}, 
and localizes into a LC 11d
black hole phase for entropies $S<N$.
The line $S\sim N$ separates the 11d phases that
are localized on the M-theory circle (whose coordinate size is $\r11$)
from those that are delocalized, uniformly across the diagram
~\cite{BFKS1,BFKS3,HORMART,BFKS2}.
The 11d black hole phase at small entropy
becomes smeared across the $T^p$ when
the horizon size becomes smaller than the torus scale $V$;
we denote generally such smeared phases by an overline
(in this case $\overline{11d}$).  This (de)localization
transition of the horizon on the compact space extends
above the $S\sim N$ transition, separating the black
Dp brane phase from the black $D0$ brane phase
\footnote{Initially, the $D0$ brane phase becomes smeared 
to $\overline{D0}$; as the entropy increases, the effective
geometry of the latter patch
becomes substringy at the horizon, and one should
T-dualize into the black Dp brane patch.  Both the
$\overline{D0}$ and $Dp$ patches have the same equation
of state, since they are related by a symmetry transformation
of the theory; they are different patches of the same phase.}.
Susskind \cite{SUSSGW} has argued that, on the SYM side,
one should regard this localization transition as an analogue
of the Gross-Witten large N transition \cite{GROSSWITTEN}.
The localization transition line runs into the correspondence
curve separating the SYM gas phase from the geometrical
phases at $S\sim N^2$.  Thus
as we move to the left (decreasing $V$, \ie\ increasing bare
SYM coupling) at high entropy $S> N^2$, the 
SYM gas phase reaches a correspondence point;
on the other side of the transition is the phase of
black Dp-branes.
A further common feature of the diagrams is a `self-duality'
point at $V\sim 1$ and $S\sim N^{\frac{8-p}{7-p}}$,
where a number of U-duality curves meet.

In contrast, the structure of the phase diagrams for $V<1$
depends very much on the specific case at hand.
Compactifications on $T^p$, $p=1,2,3$, were analyzed in \cite{MSSYM123};
we now describe the specifics of this region for $p=4,5,6$.

Figure~\ref{SYM4fig} is the phase diagram of $T^4$ compactification.
There are six different phases, several of which -- the 11d 
and $\overline{11d}$ black
hole, black $D0$ and $Dp$ brane, and SYM gas phases -- 
were discussed above.  In a slight shift of emphasis,
we have relabeled the black $D4$ brane phase as the black $M5$ brane
phase, since its description in terms of the latter
object extends to the region $V<1$ (in fact, even for a patch of $V>1$
the D4 brane becomes strongly coupled and must be lifted to
M-theory).  The appropriate dual non-gravitational description
involves the six-dimensional $(2,0)$ field theory on $T^4\times S^1$,
where the last factor is the M-theory circle;
the scale of Kaluza-Klein excitations given by the size of this circle 
(times the number of branes) sets the
transition point between the $(2,0)$ and SYM descriptions.
This $M5$ phase consists of six patches
that we cycle through via duality transformations
required to maintain a valid low-energy description. 
The energy per entropy increases toward the left and toward higher entropies;
this is to be contrasted with the cases analyzed in~\cite{MSSYM123} where the
IR limit appears toward the left of the diagrams.
This behavior is a consequence of the reversal of the direction
of RG flow between $p<3$ and $p>3$.
As we continue to the left and/or down on the figure
at small volume $V<1$, the $T^4$ is small while the M-theory circle
remains large; eventually one reduces to string theory along
the cycles of the $T^4$, and the M5-brane dualizes into
a string.  Somewhat further in this direction, we encounter
a Horowitz-Polchinski correspondence curve, and a transition to a phase 
consisting of a Matrix String~\cite{MOTL,DVV,BSMAT} 
with effective string tension set by
the adjacent geometries. Using Maldacena's conjecture, we thus
validate earlier suggestions to describe Matrix strings using the
$(2,0)$ theory~\cite{BRS,SEIBLITTLE,WHYSEIB}. 
This matrix string phase has a correspondence curve also for low entropies,
now with respect to a phase of smeared LC M theory black holes 
(or equivalently boosted IIB holes). 

Figure~\ref{SYM4fig} is trivially modified to give the phase
diagram of the $(2,0)$ theory on $T^4/Z_2\times S^1$. 
The additional structure does not affect
the critical behavior of the diagram. The change appears in the chain of
dualities we perform on the dotted lines of the diagram. 
The orbifold quotient metamorphoses into world-sheet parity, 
and the fundamental string patch (labeled $\overline{F1}$)
becomes that of the Heterotic string. The emerging Matrix
string phase at the correspondence point is then that of a Heterotic
theory. We thus confirm the suggestion~\cite{BR,DIACGOMISMS}
to describe Heterotic Matrix
strings via the $(2,0)$ theory on $T^4/Z_2\times S^1$. One can also
propose to extend the dual theory of an intermediate state obtained in the
chain of dualities between the $M5$ and the $\overline{F1}$ patches
into the Matrix string regime; we then have Heterotic Matrix strings
encoded in the $O(N)$ theory of type I D strings,
as suggested in~\cite{BANKSMOTL,LOWE,HORAVA}.
Similar statements can be made about matrix theory orbifolds/orientifolds
in other dimensions.

\begin{figure}[p]
\epsfxsize=12cm \centerline{\leavevmode \epsfbox{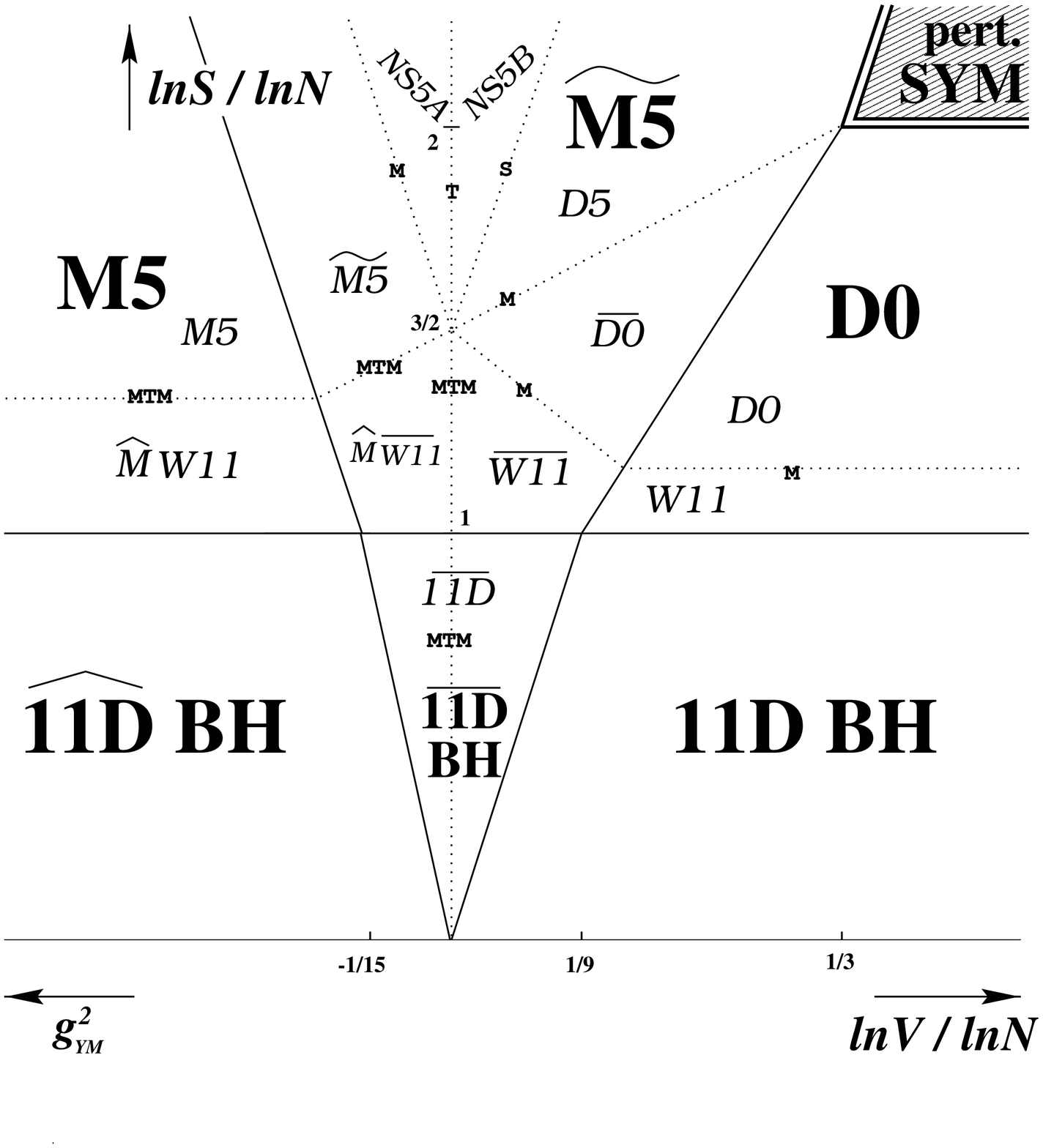}}
\caption{{\small\sl Phase diagram of `little string' theory on $T^5$.
The labeling is as in Figure~\ref{SYM5fig}.
$D0$: black D0 geometry;
$W11$: black 11D wave geometry;
$11DBH$: 11D LC black hole;
$\overline{D0}$: black smeared D0 geometry;
$\overline{W11}$: black smeared 11D wave geometry;
$\overline{11D}BH$: 11D smeared LC black hole;
$D5$: black D5 geometry;
$NS5B$: black five branes in IIB theory;
$NS5A$: black five branes in IIA theory;
$M5$: black M5 brane geometry;
$\wtilde{M5}$: black smeared M5 brane geometry;
$\what{M}\overline{W11}$: black smeared wave geometry in $\what{M}$ theory;
$\what{M}W11$: black smeared wave geometry in the $\what{M}$ theory;
$\what{11D}BH$: smeared boosted black holes in the $\what{M}$ theory.
}}
\label{SYM5fig}
\end{figure}

The thermodynamic phase diagram of fivebranes 
(sometimes called the theory of `little strings'
~\cite{DVV5D,SEIBLITTLE,DVVM5})
on $T^5$ is shown in Figure~\ref{SYM5fig}. We have a total of seven 
distinct phases.  We again shift the notation somewhat,
relabeling the black $D5$ phase as a black $\wtilde{M5}$ phase,
since the latter extends the validity of the description to $V<1$
\footnote{The tilde is meant to distinguish this eleven-dimensional phase
(where the M-circle is transverse to the five-branes) from the
eleven dimensional LC phase on the lower right, 
whose M-circle has a different origin.}.
The equation of state of this high-entropy regime is 
\bb
  S\sim E N^{\frac12}\left(\frac{\lp^2}{\r11}\right) V^{-\frac52}\ ,
\ee
characteristic of a string in its Hagedorn phase.
The temperature determines the tension of the effective string.
We have a patch of black NS5 branes 
in the middle of the diagram. They appear near the $V\sim 1$
line, at which point a T duality transformation
exchanges five branes in IIA and IIB theories. 
The IIB $NS5$ patch connects to a $D5$ brane patch via S-duality.
The IIA $NS5$ patch lifts to a patch of $M5$ branes on $T^5\times S^1$
at strong coupling on the left.  The extra circle is the M-circle
transverse to the wrapped $M5$-branes; the horizon undergoes
a localization transition on this
circle at lower entropy and/or smaller $V$ to a phase
whose equation of state is that of a 5+1d gas.
It is interesting that the Hagedorn transition is seen here
as a localization/delocalization transition in the black geometry.
Yet further in this direction,
the system localizes at $N\sim S$ to a 
a dual LC $\what M$ theory on a $T^4\times S^1\times S^1$;
here the horizon is smeared along the square $T^4$,
localized along both $S^1$ factors, and carrying momentum
along the last $S^1$. 
This $\what{M}$ phase on the lower left is U-dual to
the LC M-theory on the lower right.

\begin{figure}[p]
\epsfxsize=14cm \centerline{\leavevmode \epsfbox{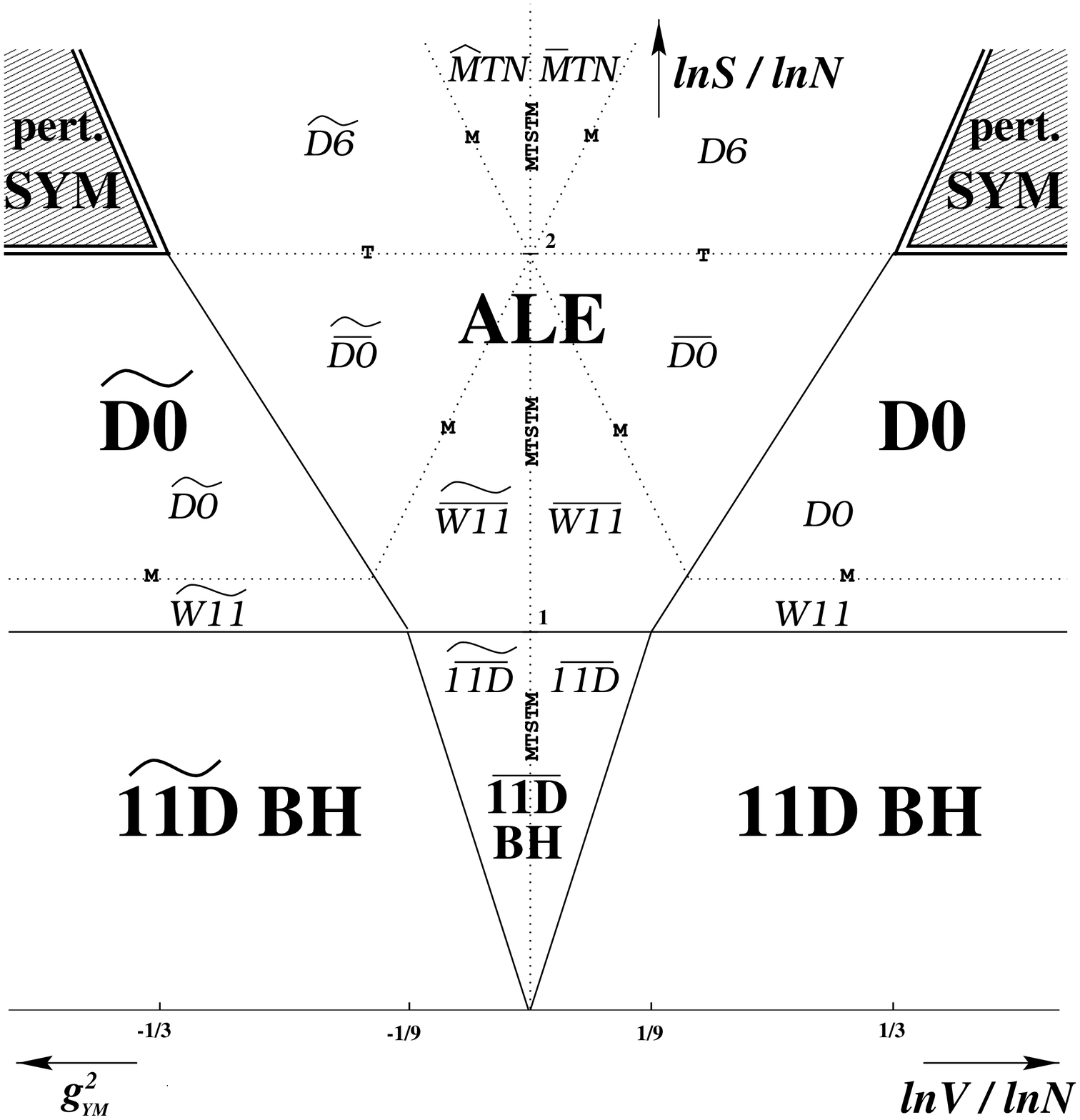}}
\caption{{\small\sl Phase diagram of the $D6$ system.
$S$ is entropy, $V=R/\lp$ is the size of a cycle on the $T^6$
of the LC M theory, and $N$ is longitudinal momentum.
The dotted lines are symmetry transformations:
M: M lift or reduction; T: T duality; S: S duality.
The solid lines are phase transition curves. Double solid lines
denote correspondence curves.
The label dictionary is as follows:
$\overline{M}TN$,$\what{M}TN$: black Taub-NUT geometry;
$D6$,$\wtilde{D6}$: black D6 geometry;
$D0$,$\wtilde{D0}$: black D0 geometry;
$W11$,$\wtilde{W11}$: black 11D wave geometry;
$11DBH$,$\wtilde{11D}BH$: 11D LC black hole;
$\overline{D0}$,$\wtilde{\overline{D0}}$: black smeared D0 geometry;
$\overline{W11}$,$\wtilde{\overline{W11}}$: black smeared 11D wave geometry;
$\overline{11D}BH$,$\wtilde{\overline{11D}}BH$: 11D smeared LC black hole.
}}
\label{SYM6fig}
\end{figure}

The $D6$ phase diagram has two important features (see Figure~\ref{SYM6fig}).
First of all, the Maldacena limit keeps fixed the Planck scale
$\tlp\sim\frac{\lp^2}{\r11}V^{-2}$ of the high-entropy
black Taub-NUT geometry\footnote{In the Maldacena limit, the near horizon
geometry is that of an ALE space with $A_{N-1}$ singularity.}~\cite{MALDA2}.
Thus, gravity does not decouple,
and the limit does not
lead to a non-gravitational dual system that would
serve as the definition of M-theory in such a spacetime.
A symptom of this lack of decoupling of gravity 
is the negative specific heat $S\propto E^{3/2}$ 
of the high-entropy equation of state.  This property
is related to the breakdown of the usual UV-IR
correspondence of Maldacena duality~\cite{SUSSWIT,PEETPOLCH}.
The energy-radius relation of~\cite{PEETPOLCH}
determined by an analysis of the scalar wave equation in the
relevant supergravity background, is in fact the
relation between the horizon radius and the Hawking temperature
of the associated black geometry; thus, 
for $p=6$ {\it decreasing energy}
of the Hawking quanta is correlated to {\it increasing radius}
of the horizon, as a consequence of the negative specific heat.
This is to be contrasted with the situation for $p<5$, where
the positive specific heat means increasing horizon radius
correlates to increasing temperature; and $p=5$, where the
Hawking temperature is independent of the horizon radius
in the high-entropy regime.
Now, temperature in any dual description must be the same
as in the supergravity description.  For $p<5$, the dual is
a field theory; high temperature means UV physics dominates
the typical interactions, leading to the UV-IR correspondence.
For $p=5$, the dual is a `little string' theory; the temperature
is unrelated to the horizon radius (and thus the total energy)
on the gravity side, and unrelated to short-distance physics
in the dual `little string' theory (since high-energy collisions
of strings do not probe short distances).  Hence the UV-IR
correspondence already breaks down at this point.  For $p=6$,
there is nothing to say -- large radius (large total energy)
corresponds to low temperature of probes (Hawking quanta);
and any dual description could not have high energy/temperature
related to short distance physics, since it is a theory
that contains gravity (so high energy makes big black holes).

A second key feature is the duality symmetry 
(\cf~\cite{SENTP}) $V\rightarrow V^{-1}$
of the diagram relating the $V<1$ structure to that discussed
above for $V>1$.  Note that this duality symmetry inverts
the $T^6$ volume as measured in {\it Planck} units
rather than string units.  The duality interchanges momentum
modes with fivebrane wrapping modes, while leaving membrane
wrapping modes fixed; in other words, the dual space is
that seen by the $M5$ brane.  It is possible that this
symmetry extends to any Calabi-Yau compactification of M-theory,
since the volume of the \CY\ sits in a universal hypermultiplet
whose moduli space appears to be $SU(2,1)/U(2)$
~\cite{STROMUH};
if the discrete identifications involve the appropriate
element of $SU(2,1;\Z)$, there will be a dual \CY\ of roughly
the inverse size seen by M-theory fivebranes wrapping the original \CY.

The thermodynamic perspective also sheds light on a proposal
of Kachru, Lawrence, and Silverstein~\cite{KLS}
for a definition of matrix theory
compactifications on \CY\ spaces.
Generically, string theory on a \CY\ space does not have a T-duality
that inverts its volume in {\it string} units.
Rather, these authors suggest that the appropriate duality to consider,
analogous to the T-duality transformation 
used by Sen-Seiberg for torus compactifications,
is the mirror symmetry transformation.  This transformation
relates $D0$-branes in IIA theory on a given \CY\
to $D3$-branes wrapping a special Lagrangian submanifold
of the IIB mirror~\cite{SYZ};
locally, the \CY\ looks like a $T^3$ fibered over an $S^3$ base,
and mirror symmetry is T-duality on the fiber.
Thus, it is proposed that some sort of 3+1 gauge dynamics might
yield an appropriate underlying description.
Consider the phase diagram that should arise.
At low entropy, one has the 11d black hole phase.
As the entropy increases at fixed 
but not large \CY\ coordinate size $V$,
one finds the horizon smears over the \CY\ and eventually
one reaches the $\overline{D0}$ patch of smeared black $D0$ branes.
The proper size of the \CY\ at the horizon 
in string units is decreasing along this path; eventually one
reaches the curve along which one should perform the duality
transformation, in this case mirror symmetry.
Naively, in the mirror, as the entropy increases further, the $T^3$
wrapped by the $D3$-branes is increasing in size, while
the base $S^3$ continues to shrink; the high-entropy phase
would seem to be described by $D3$-branes on the special
Lagrangian cycle of the mirror \CY\ near a conifold singularity.
However, the duality transformation will not change the equation
of state, since the $\overline{D0}$ patch and everything
above it are related by symmetries of the theory.
The only thing that could change this conclusion is a further
phase transition, but there is no candidate.  We conclude
that the high-entropy phase is again one with negative
specific heat, and thus cannot be that of a field theory
\footnote{Note that one could also imagine performing the same
duality sequence to describe matrix theory on K3 in terms of
two-branes on the torus fiber of a near-degenerate mirror K3.
In this case one knows that this description is related
to the five-brane description given above by duality,
hence indeed has a 5+1d equation of state at high entropy,
rather than a 2+1d equation of state.}.

\subsection{The D1-D5 system}

\begin{figure}[p]
\epsfxsize=11cm \centerline{\leavevmode \epsfbox{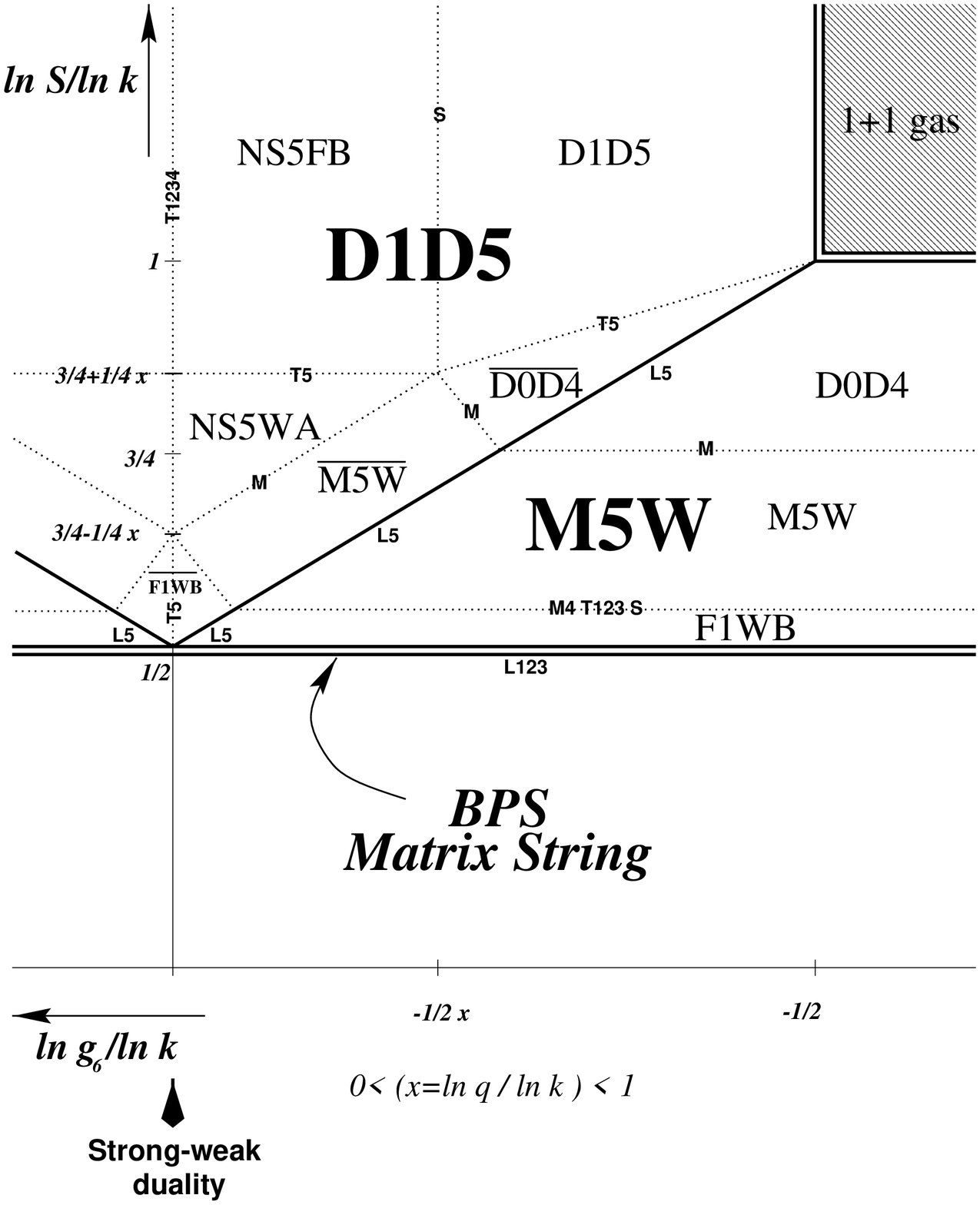}}
\caption{{\small\sl Thermodynamic phase diagram of `little strings'
wound on the $S^1$ of $T^4\times S^1$,
with $Q_1$ units of winding and $Q_5$ five branes. $k\equiv Q_1 Q_5$
and $1<q\equiv Q_1/Q_5<k$. $g_6$ is the six dimensional string coupling
of the $D1D5$ phase.
The label dictionary is as follows:
$D1D5$: black $D1D5$ geometry;
$NS5FB$: black NS5 geometry with fundamental strings in IIB theory;
$D0D4$: black D0D4 geometry;
$\overline{D0D4}$: black smeared D0D4 geometry;
$M5W$: black boosted M5 brane geometry;
$\overline{M5W}$: black smeared boosted M5 brane geometry;
$NS5WA$: black boosted NS5 branes in IIA theory;
$F1WB$: black boosted fundamental strings in IIB theory;
$\overline{F1WB}$: 
black smeared and boosted fundamental strings in IIB theory;
L: localization transitions.
}}
\label{D1D5fig}
\end{figure}

As a further example of our methods, we have examined the D1-D5 system
on $T^4\times S^1$, which as we mentioned above can be considered
as the `little string' theory of $Q_5$ fivebranes, with $Q_1$ units of
string winding along the $S^1$. 
Figure~\ref{D1D5fig} shows the thermodynamic phase diagram.
In the Maldacena limit,
this theory is a representation of the algebra of $\NN=(4,4)$
superconformal transformations in 
1+1d~\cite{MALDA1,STROMBHMICRO,BROWNHEN,MALDASTROMADS3,MARTINECADSMM,GKS}.
We have defined $k\equiv Q_1 Q_5$ and $q\equiv Q_1/Q_5$.  
We keep $k$ fixed, but $q$ may be viewed as a variable
ranging between $1<q<k$, thus moving some of the dotted curves
of duality transformation, but not altering phase transition curves.
For $q\sim 1$, we can exchange
the roles of $Q_1$ and $Q_5$ via duality transformations across the
diagram; the structure is unchanged. 
The other limit, $q=k$, is the $Q_5=1$ bound.
The vertical axis on the diagram is again the entropy, 
while the horizontal axis is the six-dimensional string
coupling $g_6\equiv g_s/\sqrt{v}$ of the $D1D5$ patch,
where $v=V_4/\alp^2$ is the volume of the $T^4$ in string units
(equivalently $g_6^{-2}$ is the 
volume of the $T^4$ in appropriate string units of the NS5FB phase). 
The phase diagram has a symmetry $g_6\rightarrow 1/g_6$
(inversion of the torus in the NS5FB phase); 
this is the T-duality symmetry of the little string theory.
From the perspective of the $D1D5$ patch,
we can consider the entire phase diagram as that of the
1+1d conformal theory that arises in the IR of this gauge
theory, which is conjectured to be dual to the near-horizon 
geometry $AdS_3\times S^3\times T^4$ of the $D1D5$ system.
In this patch, the D strings are wrapped
on a cycle of size $R_5$.  This parameter is absent from the scaling 
relations of all curves because of conformal symmetry.
Analogous to the singly-charged brane systems we have been
discussing, at high entropies there is a `1+1d gas' phase at
small $g_6$ (large $V_4$), which passes across a correspondence curve
to the black brane phase as the coupling increases.
Being determined by conformal symmetry and quantization of the
central charge, the equation of state does not change across
this `phase transition'.
Starting in the `1+1d gas' phase and decreasing the entropy,
$S\sim k$ corresponds to the point 
where the thermal wavelength in the 1+1d conformal theory
becomes of order the size of the box $R_5$. 
This is again a Horowitz-Polchinski correspondence
curve from the side of lower entropies, 
analogous to the SYM theories at $S\sim N^2$
\footnote{There is similarly a hidden phases of zero specific
heat between the gas phase and the lower, localized phase,
as can be seen by the discontinuity in temperatures
that occurs between $S>k$ and $S<k$.}.
There is a localization transition on the $R_5$ cycle
cutting obliquely across the diagram. The localized phase
can be interpreted as that of $M5$ branes with
a large boost, thus connecting with 
the proposal of~\cite{ABKSS}
for a matrix theory of this system.
The lower boundary of this phase occurs at
entropies of order $S\sim \sqrt{k}$,
where a BPS Matrix string phase emerges and the diagram is
truncated at finite entropy.   
We find agreement with Vafa's argument~\cite{VAFAHAGE}
that the BPS spectrum in the R sector of the D1-D5 system 
is that of fundamental IIB strings carrying
winding and momentum 
(sometimes called Dabholkar-Harvey states~\cite{DABHOLHARV}).
Similarly, chasing through the sequence of dualities for
the D1-D5 system on $K3\times S^1$, one finds the BPS spectrum
consists of Dabholkar-Harvey states of the heterotic string.

The character of the phase diagram is different at the extreme limits
$q\sim 1$ (\ie\ $Q_1\sim Q_5$) and $q\sim k$ (\ie\ $Q_5\ll Q_1$).
The location of the transition curves bounding the $NS5WA$ patch 
(type IIA NS five-branes with a wave as the low-energy description)
depend on the ratio $x={ln\,q}/{ln\,k}$.  For roughly
equal charges $q\sim1$,
$x\sim 0$, this patch disappears, as do the related
$\overline{M5W}$ and $M5W$ patches and the
$NS5FB$ patch of fundamental strings and IIB NS5-branes.
The D-brane description predominates the phase diagram,
except at low energies where there is a large patch describing
fundamental strings with winding and momentum.
The opposite regime, say fixed $Q_5$ and large $Q_1$ so that $x\sim 1$,
is the regime discussed by~\cite{GKS};
it is also relevant to the `DLCQ' description of the fivebrane~\cite{ABKSS}.
Indeed, the high-entropy region $S>k$ is taken over by the $NS5FB$
patch up to the correspondence curve; while in the low-entropy 
domain $S<k$, the $NS5WA$ patch expands to squeeze out the
$\overline{D0D4}$, $\overline{M5W}$, and $\overline{F1WB}$ patches,
and the localized phase is covered by the $M5W$ patch -- longitudinal
M-theory five-branes with a large boost, just what
one needs for an infinite momentum frame or DLCQ description.
We discuss the DLCQ limit in detail in section \ref{DLCQsect} below
\footnote{The relation between the Maldacena conjecture
and matrix models of M5-branes has also been considered
recently in \cite{AWATAHIRANO}.}.  

For simplicity, we have restricted the set of parameters 
we have considered in the phase diagram to the entropy and
the coupling $g_6$.  It is straightforward to see what
will happen as other moduli of the near-horizon geometry
are varied.  Consider for instance decreasing one of the $T^4$
radii keeping the total volume fixed.  At some point, the
appropriate low energy description will require T-duality
on this circle, shifting from 
$D1$-branes dissolved into $D5$-branes, to $D2$-branes
ending on $D4$-branes.  One can then chase this duality
around the diagram: The $NS5FB$ phase becomes $M2$-branes
ending on $M5$-branes; the $NS5WA$, $\overline{D0D4}$, and
$D0D4$ phases become $D1$-branes ending on $D3$-branes;
and the $\overline{M5W}$ and $M5W$ phases become those of
fundamental strings ending on $D3$ branes.  The near-extremal
$F1WB$ phase is unaffected.
One can also imagine replacing the $T^4$ by K3.  Moving around
the K3 moduli space, when a two-cycle becomes small,
a $D3$-brane wrapping the vanishing cycle becomes light;
one should consider making a duality transformation 
that turns $Q_1$ or $Q_5$ into the wrapping number on this cycle.

Thus the D1-D5 system appears to have a remarkably varied life. 
On the one hand, it can describe low-energy supergravity
on a 6d space, namely $AdS_3\times S^3$; the common
coordinate of the branes is the angle coordinate on $AdS_3$.
This space parametrizes physics of the Coulomb branch of the
gauge theory.  On the other hand, the same system describes 
the `decoupled' dynamics of the five-brane, another 6d system
\footnote{Seven-dimensional, if we include the circle
transverse to the $M5$-brane.} --
except that the spatial coordinates are now $T^4\times S^1$,
with the $T^4$ apparently related to the physics of the Higgs
branch of the gauge theory, and the $S^1$ the dimension common 
to the branes.  In the Maldacena limit, the theory is
a representation of the 1+1d superconformal group;
in the DLCQ limit, it describes light-cone $M5$-branes.

The careful reader will have noted that we have refrained
from characterizing the nongravitational dual 
of the D1-D5 geometry as a 1+1d field theory
\footnote{The following remarks reflect ongoing discussions
of the first author with D. Kutasov and F. Larsen.  In particular,
it was D. Kutasov that raised the question of whether the
dual object is a field theory.}.
The standard candidate for this dual is the 1+1d 
conformal field theory (CFT) on $Sym^k(T^4)$ (or K3).
This CFT is supposed to provide a description of nonperturbative
string theory on $AdS_3\times S^3\times T^4$ (or K3).
Indeed, it captures the high-entropy thermodynamics
\cite{STROMVAFA} as well as the BPS spectrum 
\cite{VAFAHAGE,MALDASTROMADS3,DEBOER}.
However, the near horizon geometry appears to put
the CFT at a singular point in its moduli space \cite{WITTENHIGGS,DIJKGRAAF};
also, there appears to be a mismatch 
in the level of the $U(1)^4$ affine algebra of Noether charges
acting on the $T^4$ \cite{KLL}.  A basic problem also
arises in the phase diagram of Figure \ref{D1D5fig}. 
In the high-entropy phase,
S-duality connects the $D1$-$D5$ patch to the $NS5$-$F1B$ patch
as one moves to stronger coupling.
It is straightforward to check that, crossing the boundary 
$g_6\sim q^{-1/2}$, the energy scale of a $D1$-brane
wrapping the torus $T^4$ becomes less than that of a fundamental string;
the appropriate effective description is the S-dual one.
In fact there has to be an entire decuplet of strings transforming
under the $O(5,5;Z)$ U-duality group; 
the proper low-energy description favors 
one pair of these, electrically and magnetically charged
under one of the five six-dimensional $B$-fields (the subgroup
of U-duality fixing the description is $O(5,4;Z)$)%
\footnote{There are BPS charges corresponding to these objects wrapping
$T^5$, which are central charges in the 10d supersymmetry algebra.
Just as in the case of the transverse five-brane
in Matrix theory \cite{BSS},
these charges decouple from the supersymmetry algebra
in the Maldacena limit; nevertheless the objects remain as
finite energy excitations carrying conserved charges.}.
The problem is that the objects 
carrying these charges, which are the lightest objects
in the theory at intermediate coupling, are not apparent in the
$Sym^k(T^4)$ symmetric orbifold anywhere on its moduli space.
Similarly, in the D1-D5 system on K3 there should be 
a full $O(5,21)$ 26-plet of strings; in this case, tensionless strings
corresponding to wrapped $D3$-branes
arise when a 2-cycle on the K3 degenerates, and 
are essential in order to regularize
the singularities in the effective description.
The 1+1d CFT on (symmetric products of) K3 is simply singular, 
and does not contain the objects which are needed.  
These objects are carried, however,
as fluxes on the five-brane one starts with;
the energy cost of these excitations simply becomes small
at the relevant points in moduli space, suggesting
that the 5+1d string-theoretic
character of the dynamics does not fully decouple
in the Maldacena limit.  Similarly, one might expect
that lower dimensional examples of the Maldacena conjecture
(\eg\ those involving $AdS_2$ or $AdS_3$)
are not fully captured by quantum mechanics or more elaborate 1+1d
field theories.
As mentioned above, it is known that the background fields
of the near-horizon limit of the D1-D5 system 
correspond in the symmetric orbifold CFT to turning off the
CFT resolution of the $\Z_2$ singularities of $Sym^k(T^4)$.
It may be that branes wrapping these vanishing cycles are again
the needed ingredient for a well-defined description
at these points of moduli space.

\subsection{\label{specflowintro}Spectral flow and angular momentum}

The 1+1d $\NN=(4,4)$ superconformal algebra has two canonical realizations,
depending on whether one chooses antiperiodic (NS) or
periodic (R) boundary conditions on the fermionic generators.
The spacetime geometry in the Maldacena limit
of the D1-D5 system is $AdS_3\times S^3\times \MM_4$.
2+1d supergravity on asymptotically locally $AdS_3$ spacetimes
carries a realization of this superconformal
algebra~\cite{BROWNHEN,STROMBHMICRO}; 
being a subgroup of the (super)diffeomorphism group, the symmetry
extends to the full string theory~\cite{GKS}.
$AdS_3$ itself 
is the vacuum state, and lives in the NS sector
since the Killing spinors are antiperiodic; hence
low-energy supergravity about this vacuum is described
by NS sector representation theory.  The R sector is what
one naively discovers as the near-horizon limit of
D1-D5 bound states on $\MM_4\times S^1$, since the supercharges
are periodic on $S^1$.

A similar situation occurs, for example, in $D3$-brane
gauge theory.  The gauge theory on $S^3$ describes supergravity
on $AdS_5\times S^5$ in `global coordinates'~\cite{HOROOGURI},
where time translation is 
generated by the dilation operator in the conformal group.  
The gauge theory on $\IR^3$ (or $T^3$)
describes supergravity on a slice of $AdS_5\times S^5$
in `Poincar\'e coordinates' (with periodic identifications for $T^3$), 
where time translation is generated by a conformal boost operator.
The Poincar\'e slice is obtained as the 
limiting near-horizon geometry of black $D3$-branes
in the full string theory.
There is no map between gauge theory on $S^3$
and gauge theory on $T^3$.

A major difference in the D1-D5 system is that, 
since the one-dimensional sphere and torus are the same, 
the NS and R sectors can be related 
by a continuous twist of boundary conditions
known as {\it spectral flow}.  This operation
shifts conformal dimensions $(h_L,h_R)$ 
and $J_3$ charges $(j_L,j_R)$ by \cite{SCHWIMSEIB}
\bbb
  h_{L,R}^{(\eta)}& = &h_{L,R}^{(0)}-2\eta j_{L,R}^{(0)}+\eta^2 k 
	\nonumber\\
  j_{L,R}^{(\eta)}& = &j_{L,R}^{(0)}-\eta k\ .
\label{spectralflow}
\eee
Here, $J_a$ are the SU(2) chiral R-symmetry currents
of the $\NN=(4,4)$ algebra; $E=\frac12(h_L+h_R)$ is the energy; 
$P=\frac12(h_L-h_R)$ is the momentum along $x_5$.
We will restrict our attention to the $P=0$ sector.
The mode expansions of the supercurrents
(which have $j=\pm\hf$) are shifted by $\pm\eta$.
Thus spectral flow by $\eta=n+\hf$, $n\in\Z$ 
relates NS sector states to R sector states.  
Moreover, spectral flow by integral amounts $\eta\in\Z$ maps a given
sector onto itself; the spectrum maps to itself, 
but individual states are not preserved.
This, combined with the charge conjugation
symmetry $j\rightarrow -j$, means that the full spectrum of
states in the theory (for both NS and R sectors)
with $j_L=j_R=j$
is determined by, \eg, NS sector states with $0\le j\le k/2$.
This relation implies a relation between standard conventions
in the literature: $h_{(\rm Ram)} = h_{(\rm NS)}-k/4$,
and $j_{(\rm Ram)} = 0$ corresponds to $j_{(\rm NS)}=k/2$.

\begin{figure}
\epsfxsize=13cm \centerline{\leavevmode \epsfbox{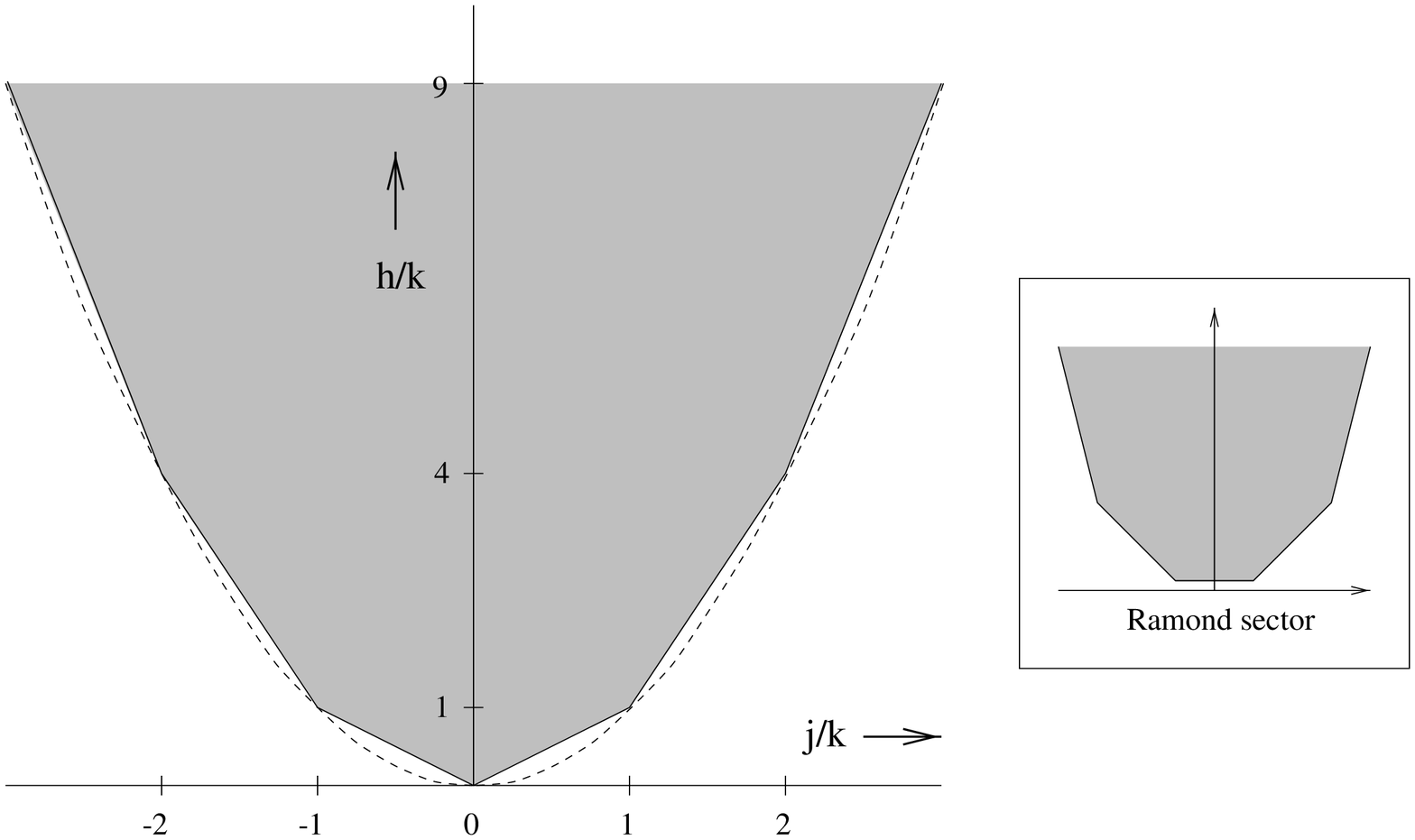}}
\caption{{\small\sl Allowed region for states belonging
to unitary representations of the (NS) superconformal algebra.
The dashed curve represents the continuous spectral flow 
$h_\eta=j_\eta^2/k$ of the point $h=j=0$.
Spectral flow slides the boundary polygon along the parabola;
a half unit of flow gives the Ramond sector (inset).
}}
\label{unitarity}
\end{figure}

In fact, there is a simple operation on the full string theory
that reduces to spectral flow in the near-horizon limit
of the D1-D5 bound state: 
It is the orbifold described by Rohm~\cite{ROHMSUSY}.
The $SU(2)_L\times SU(2)_R$ R-symmetry of the near-horizon
supersymmetry of the D1-D5 system
is inherited from the Lorentz group of the 
asymptotically flat spacetime in which it is embedded
in the original string theory.  
Thus the R-symmetry twist is nothing but the imposition of the
twisted boundary condition
\bb
  \Phi(x_5=R_5)=\exp[i4\pi\eta(J^3_L+J^3_R)]\Phi(x_5=0)\ .
\label{rohmtwist}
\ee
In the near-horizon region, the geometry is asymptotically
$AdS_3\times S^3\times\MM_4$, and the spectral flow
operation can be understood \cite{CKS} in the effective Chern-Simons
supergravity theory that arises \cite{IZQUIERDOTOWNSEND}.
There, spectral flow is implemented by coupling
the $U(1)\times U(1)$ Cartan R-symmetry currents to a source;
a shift in the energy arises due to 
the usual relation between regularization (framing) of Wilson line sources
and conformal spin in Chern-Simons theory 
\cite{WITTENKNOTS}.\footnote{Thus, 
very little of the quantum structure of gravity
is being used here.}
It is interesting that, although this twist breaks supersymmetry
in the full theory, anti-de Sitter supersymmetry is restored
in the near-horizon region; $\eta=\hf$ maps the R sector
of the wrapped brane system to the NS sector, with the R ground
state of maximal charge mapping to the NS vacuum.

Unitarity implies that any allowed highest weight
representation of the superconformal group must have $h \ge |j|$
\cite{BFK}.  
Spectral flow then implies that allowed states
must lie inside the shaded region of the $(h,j)$ plane in
Figure~\ref{unitarity}.\footnote{These 
curves are slightly different from the unitarity
boundaries of~\cite{BFK,KENTRIGGS}
since we are only asking that a state is the spectral flow
of some state in an allowed representation, rather than that
it is an allowed superconformal {\it highest} weight.}
In particular, spectral flow forces a cutoff on the
spectrum of BPS supergravity states (regardless of whether
they are single- or multi-particle configurations)
at $j=k$; as is easily seen from the figure,
states on the line $h=j$ beyond this point lie outside
the allowed region (since they would have to flow from
states that violate the BPS bound).
This feature was termed the `stringy exclusion principle'
in~\cite{MALDASTROMADS3}; we see that it depends only on some
rather mild assumptions about the quantization of
Chern-Simons supergravity (\ie, the global structure of
the class of geometries under consideration).
All such restrictions disappear 
in the classical $k\rightarrow\infty$ limit.

\begin{figure}
\epsfxsize=10.5cm \centerline{\leavevmode \epsfbox{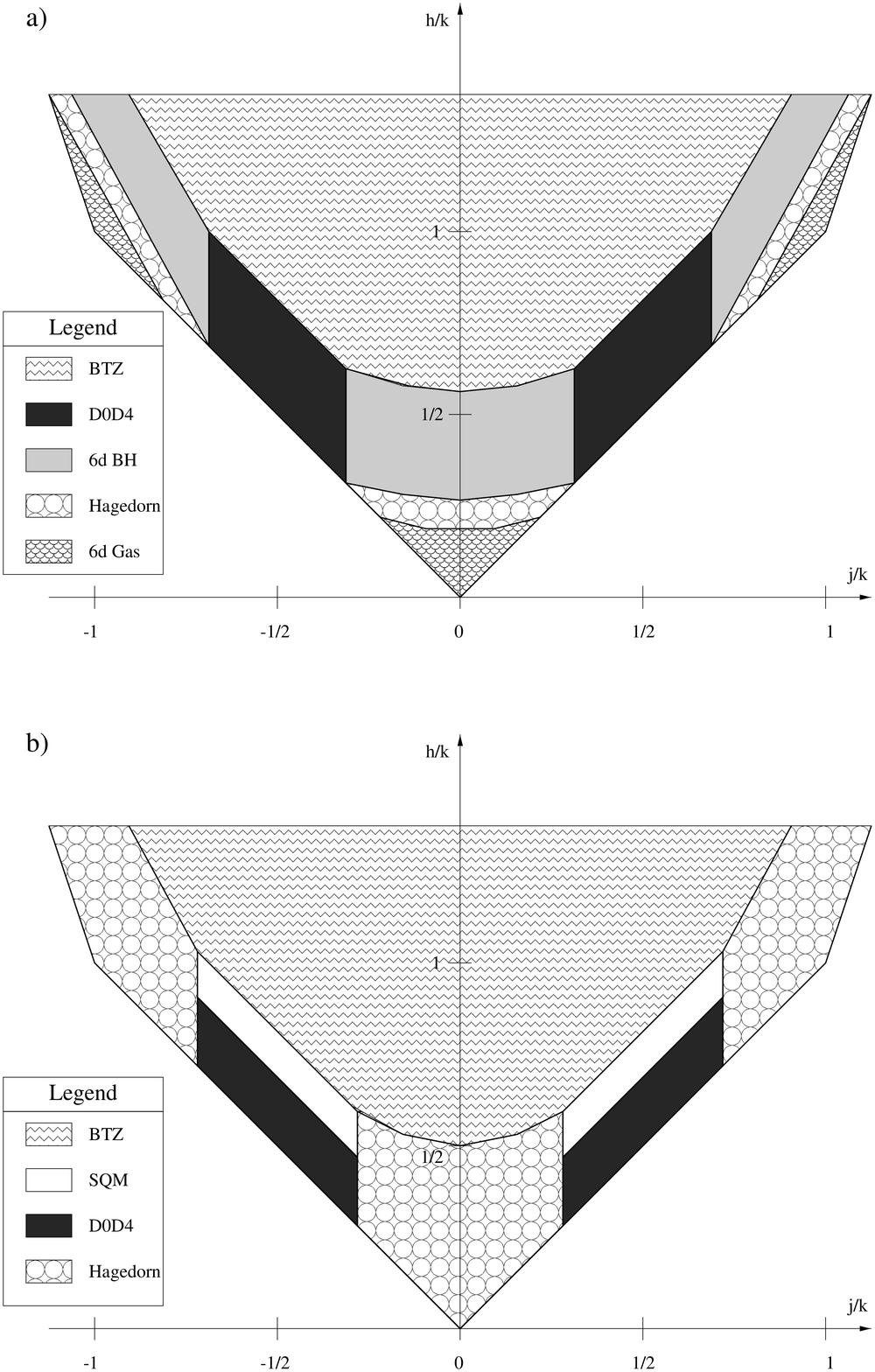}}
\caption{{\small\sl Qualitative phase diagrams for
the D1-D5 system as a function of energy and angular momentum:
a) for coupling $\geff\equiv g_6^2 k>1$,
where $g_6$ is the six dimensional string coupling;
b) for coupling $\geff<1$. 
SQM stands for Super Quantum Mechanics~\cite{BKR}, a 
phase corresponding to energy indepedent entropy $S\sim k$.
}}
\label{hversusj}
\end{figure}

Spectral flow determines the density
of states -- at high entropy and
far from the boundary of the allowed region --
in terms of the Cardy formula~\cite{CARDY,KUTSEIB} for zero charge
\bb\label{spineos}
  S=2\pi\sqrt{k(h_L-\coeff14 k)-j_L^2} 
	+ 2\pi\sqrt{k(h_R-\coeff14 k)-j_R^2}\ ,
\label{denstates}
\ee
which is precisely the density of states for D1-D5
black holes with angular momentum (remembering the
shift in conventions).
The expression must be invariant under spectral flow,
when the thermal wavelength is much smaller than
the size of the system, because the fermion boundary conditions
are irrelevant.
Near the boundary of the allowed region, the density
of states will differ from this expression.

A qualitative sketch of the phase diagram as a function of
energy and angular momentum is given in Figure~\ref{hversusj}.
The locations of the phase boundaries are not 
precisely determined,\footnote{Since 
we are now considering finite $k$, the boundaries
between phases are not sharp anyway; they are crossover
transitions rather than singularities in derivatives of the free energy.}
since we only accurately know the phase structure
in the vicinity of the NS ($j=0$) and R ($j=k/2$) sectors.
The R sector structure is that of Section \ref{litstr}, 
and outlined in the previous section;
the NS phase structure was discussed in \cite{BDHM}:
There is a `supergravity gas' phase (\ie\
the predominant states are dressed
Fock space states of low-energy supergravity) about the AdS vacuum;
at somewhat higher energy the entropy is dominated by a
long string phase; then the string undergoes a correspondence
transition to a 5+1d Schwarzschild black hole (\ie\ localized
on $AdS_3\times S^3$ and smeared on $\MM_4$); and finally,
at high energy the BTZ black hole phase with equation of state
\pref{spineos} takes over, as the 5+1d black hole delocalizes on $S^3$.
As a function of angular momentum, there are then phase boundaries
where the NS and R structures abut one another.
More details may be found in section \ref{specflowsec}.

\section{The details for the phase diagrams}

The details of our results can be found in the coming sections. 
The $D4$, the $D4$ on an orbifold,
$D5$, $D6$ and $D1D5$ systems are analyzed in detail in sections
2.1, 2.2, 2.3, 2.4 and 2.5 respectively. 
The discussion about spectral flow
and angular momentum can be found in Section 2.6.

\subsection{\label{6dqft}The $(2,0)$ theory on $T^4\times S^1$}

\sect{The M5 phase} 
Our starting point will be $D4$ branes wrapped on the $T^4$
T-dual to the matrix theory description. 
This phase consists of six geometrical patches and
is described by the equation of state
\bb\label{M5eos}
E\sim \frac{\r11}{\lp^2} V^{8/5} S^{6/5} N^{-3/5}\ ,
\ee
obtained from the geometry of $N$ D4 branes.
The geometries are parameterized by the harmonic functions
\bb
H=1+\frac{q^3}{r^3}\quad ,\qquad
h=1-\frac{r_0^3}{r^3}\ ,
\ee
with
\bb
r_0^5\sim \frac{S^2}{N} \lp^5 V^{-4}\quad ,\qquad 
q^3\sim \frac{N}{V^4} \frac{\lp^5}{\r11^2}\ .
\ee
We next describe the six patches of this phase.

\noindent
{\it The black D4 brane geometry (D4)} is given by the metric and dilaton
\bbb\label{D4geom}
ds_{10}^2&=&H^{-1/2} \lk( -h dt^2+dy_{(4)}^2\re) + H^{1/2} \lk( h^{-1} dr^2+
r^2d\Omega_4^2\re)\nonumber\\ 
e^\phi&=&H^{-1/4}\ ,
\eee
We are using the convention that
the asymptotic values of the dilaton
are absorbed into the gravitational coupling.
The parameters of this geometry are related to the moduli of the 
Light Cone M theory introduced above as follows:
\bb
\gs=\lk(\frac{\lp}{\r11} \re)^{1/2} V^{-4}\quad , \qquad
\alp=\frac{\lp^3}{\r11}\quad , \qquad
y\approx \frac{\lp^2 V^{-1}}{\r11}\ ,
\ee
where in the last equation, we use the notation $\approx$ to denote the
compactification scale for the four $y$ coordinates, all assumed 
equal in size. 
This geometry is subject to the following restrictions:

\begin{itemize}

\item The Horowitz-Polchinski correspondence principle requires
\bb
S> V^{12} N^{-2}\ .
\ee
Otherwise, we connect to a phase described by perturbative 4+1d SYM.

\item Requiring that the $y$'s are bigger than the string scale yields
\bb\label{D0D4}
S> V^2 N^{4/3}\ .
\ee
Otherwise, we T-dualize into the geometry of $N$ smeared black D0 branes.

\item Requiring small coupling at the horizon yields
\bb\label{M5D4}
S< V^{12} N^{4/3}\ .
\ee
Beyond this point, we describe the vacuum via the geometry of black M5 branes.
\end{itemize}

\noindent
The T duality transformation yielding the geometry of
smeared D0 branes beyond~\pref{D0D4} leads us for $V>1$
onto a phase structure identical to the ones
encountered in the three cases studied in~\cite{MSSYM123}.
We will therefore be brief in the description of the right half of the
the phase diagram;
a complete discussion can be found in the cited paper. 
We sketch
quickly the scaling of the various transition curves encountered along this
chain in the M5 phase.

\noindent
{\it The smeared black D0 geometry ($\overline{D0}$)}
localizes on the $T^4$ for
\bb\label{D0loc}
S< V^{9/2} N^{1/2}\ ,
\ee
into a phase of localized black D0 branes,
and gets M lifted to 
{\it smeared M theory black waves ($\overline{W11}$)} at
\bb\label{W11D0}
S\sim N^{4/3} V^{-4/3}\ .
\ee
At
\bb
V\sim 1\ ,
\ee
it is seen to be necessary to
reduce this latter geometry along one of the cycles of the $T^4$
to the geometry of IIA waves, then to T dualize on the remaining $T^3$ to a 
IIB theory, and finally to S dualize to the geometry of black IIB waves, to be
discussed below. The $\overline{W11}$ geometry furthermore collapses 
at
\bb\label{ssimn}
S\sim N
\ee
into the phase described by the geometry of
Light Cone M theory black holes smeared on the $T^4$.

\noindent
{\it The black M5 geometry ($M5$)} is obtained from the $D4$ geometry we started with
by lifting it, at strong couplings, to an 
$\wtilde{M}$ theory. It is described
by the metric
\bb
ds_{11}^2=H^{-1/3} \lk(dx_{11}^2+dy_4^2-h dt^2\re) + H^{2/3} \lk( h^{-1} dr^2
+r^2 d\Omega_4^2\re)\ ,
\ee
and the $\wtilde{M}$ theory is parameterized
\bb
\tlp^3= \lp \lk( \frac{\lp^2}{\r11}\re)^2 V^{-4}\quad , \qquad
\tr11= \frac{\lp^2}{\r11} V^{-4}\quad , \qquad
y_4 \approx \frac{\lp^2}{\r11} V^{-1}\ .
\ee
This geometry is subject to the following restrictions:

\begin{itemize}

\item Requiring that the curvature at the horizon is greater than the 
Planck scale $\tlp$ yields
\bb
N>1\ ,
\ee
\ie\ there is no 
dual geometrical description for $N\sim 1$. Whatever
the string theoretical description of a few M5 branes
is to be, it will take over the phase description beyond this point.

\item Requiring that the $T^4$, as measured at the horizon, is bigger than
the Planck scale yields the condition
\bb\label{mts}
S> V^{-3} N^{4/3}\ .
\ee
We must otherwise reduce to the geometry of $\wtilde{D4}$ in some
$\wtilde{IIA}$ theory living on $T^3\times S^1$
(where we have isolated an arbitrary one of the four circles
to be that of M-reduction to IIA).
\end{itemize}

\noindent
Reducing to $\wtilde{D4}$ branes wrapped on
$T^3\times S^1$, we find that the size of the $T^3$ as measured at the horizon
is smaller than the string scale set by $\talp$ for entropies satisfying the
reverse of~\pref{mts}. We then T dualize the $\wtilde{D4}$ branes 
to $\wtilde{D1}$
branes wrapped on $S^1$. We find than the IIB string coupling measured
at the horizon is bigger than one for the reverse of~\pref{mts}. We then
S dualize to the geometry of {\it IIB black 
fundamental strings smeared on the $T^3$ ($\overline{F1}$)}
\bbb\label{f1bar}
ds_{10}^2&=&H^{-1} \lk( dx_{11}^2 - h dt^2\re) + dy_3^2 + h^{-1} dr^2 +
r^2 d\Omega_4^2\nonumber\\
e^\phi&=&H^{-1/2}\ .
\eee
The IIB theory is parameterized by
\bb\label{F1params}
\tgs=\frac{\lp V}{\r11}\quad , \qquad
\talp=\lp^2 V^{-4}\quad , \qquad
y_3\approx \lp V^{-2}\quad , \qquad
\tr11\approx \frac{\lp^2}{\r11}V^{-4}
\ee
This geometry is the correct dual in this phase provided that:

\begin{itemize}

\item The curvature at the horizon is smaller than the string scale $\talp$
\bb\label{corrloc}
S>V^{-3} N^{1/2}\ .
\ee
Beyond this point, the stringy description is that of a highly excited
Matrix string, as we will see shortly.

\item The $T^3$ as measured at the horizon is smaller than
the transverse size of the object (set by the angular part of 
the metric); this
yields again~\pref{corrloc}. As the box size becomes bigger than
the size of the object, the system localizes on the $T^3$. 
Taking into account the changes to the geometry 
and thermodynamics as in~\cite{MSSYM123}, 
\bbb\label{localization}
dz_{(p)}^2+f^{-1} dr^2+r^2 d\Omega_d^2&\rightarrow&
f^{-1} dr^2+r^2 d\Omega_{d+p}^2\nonumber\\
r_0^3&\rightarrow& r_0^6\sim \lp^6 S V^{-15/2} N^{-3/4}\nonumber\\ 
q^3&\rightarrow& q^6\sim \frac{\lp^8}{\r11^2} N V^{-10}\ ,
\eee
we find that
the localized fundamental string has its Horowitz-Polchinski point again 
at~\pref{corrloc}. 
Furthermore, as needed for consistency with this statement,
we find that the change in the equation of state 
for this localized phase does not affect the analysis regarding
the Matrix string phase we will perform later.
Other restrictions on the localized $F1$ geometry are all seen to be satisfied
in the region of the parameter space of interest.

\item The $T^3$ as measured at the horizon must not be substringy. 
We find than the size of the torus as measured at the horizon is at the
self-dual point.

\item The size of $x_{11}$ as measured at the horizon is
greater than the string scale $\talp$
\bb
S>V^2 N^{4/3} \ .
\ee
We otherwise T dualize on $x_{11}$, along the string, and obtain the geometry
of smeared IIB black waves.

\item Localization on $x_{11}$ is of no concern, since the symmetry structure
of the metric does not allow the Gregory-LaFlamme localization
\pref{localization} (\ie, the brane is stretched
along this cycle).
\end{itemize}

\noindent
{\it The IIB smeared black wave geometry ($\overline{WB}$)} is 
the T dual on $x_{11}$ of $\overline{F1}$ (equation~\pref{f1bar})
\bbb\label{wbbar}
ds_{10}^2&=&\lk(H-1\re) \lk(dx_{11}-dt\re)^2+dx_{11}^2-dt^2
+H^{-1} \lk(1-h\re) dt^2+dy_3^2+h^{-1} dr^2 + r^2 d\Omega_4^2\nonumber\\
e^\phi&=&1\ ,
\eee
and the IIB theory is parameterized by
\bb
\tgs= V^3\quad , \qquad
\talp = \lp^2 V^{-4}\quad , \qquad
\tr11\approx \r11\quad , \qquad
y_3\approx \lp V^{-2}
\ee
The relevant restrictions are:

\begin{itemize}

\item Localization on $x_{11}$ occurs at
\bb
S\sim N\ .
\ee
The system collapses into a new phase described by the geometry of a
boosted IIB
black hole smeared on $T^3$.

\item The string coupling at the horizon becomes bigger than one at
\bb
V\sim 1\ .
\ee
We then are instructed to perform the chain of dualities $S,T_{(3)},M$,
bringing us back to the geometry of Light Cone M theory black waves 
$\overline{W11}$.

\end{itemize}

\noindent
We thus conclude the analysis of the $M5$ phase. The dual theory can
be inferred from the $M5$ patch; it is the six-dimensional
$(2,0)$ theory wrapped on $T^4\times S^1$. 
Extending the validity of this theory throughout
the phase diagram, we conclude that we can interpret it as the phase
diagram of the $(2,0)$ theory. 
We now move onto the other phases of the $(2,0)$ theory; 
we will be brief in the discussion of the right half of the diagram, since
it overlaps in content with the lower dimensional SYM cases~\cite{MSSYM123}.

\sect{The smeared IIB black hole ($\overline{10D} BH$)}
This phase is described by the equation of state
\bb\label{sm11Dhole}
E\sim  \lk(\frac{\r11}{N} \frac{1}{\lp^2}\re) V^{8/5} S^{8/5}\ .
\ee
and consists of the IIB hole obtained from the IIB wave geometry
$\overline{WB}$ at $S\sim N$,
and the smeared $11D$ LC hole obtained from the $11D$ 
wave geometry $\overline{W11}$.
Its correspondence point can be found by minimizing the Gibbs energies between
the equation of state~\pref{sm11Dhole}
and that of the Matrix string, which we perform below.
The smeared hole geometry localizes on the $T^4$ at
\bb
S \sim V^9\ ,
\ee
where the localized $11D$ LC black hole emerges.

\sect{The black D0 phase ($D0$)} 
This phase consists of the
geometries of {\it localized black D0 branes ($D0$)} and its M lift
{\it Light Cone M theory waves ($W11$)}; the two patches meet at
\bb
S\sim N^{8/7}\ .
\ee
The equation of state is
\bb
E\sim \lk(\frac{\r11}{N}\frac{1}{\lp^2}\re) S^{14/9} N^{2/9}\ ,
\ee
obtained from the $D0$ geometry.
The $W11$ patch
collapses into a Light Cone M theory black hole phase at~\pref{ssimn}.
The black D0 brane patch has its Horowitz-Polchinski correspondence
point at $S\sim N^2$. This is an interesting transition discussed in greater
detail in~\cite{MSSYM123}; on the $S-V$ phase diagram, the 4+1d perturbative
SYM phase emerges beyond this point.

\sect{The 11D black hole phase ($11D$ BH)} 
The equation of state is
given by
\bb
E\sim \frac{\r11}{\lp^2}N^{-1} S^{16/9}\ .
\ee
More details about this phase can be found in~\cite{MSSYM123,LMS,HORMART}.

\sect{The Matrix string phase} 
The $\overline{F1}$ geometry
encountered above breaks down via the Horowitz-Polchinski principle of
correspondence at~\pref{corrloc}. 
The emerging phase is that of a Matrix string. This can be verified
as follows: using the 
string scale given in this geometry~\pref{F1params}, 
we can write down the equation 
of state of the Matrix string phase
\bb\label{matstringeos}
E\sim \frac{\r11}{\lp^2} N^{-1} S^2 V^4\ .
\ee
Matching this energy with that of the M5 phase~\pref{M5eos}, (or 
that of the localized $F1$ geometry), 
yields~\pref{corrloc}.
Similarly, we can match the equation of states~\pref{matstringeos} and
that of the IIB hole~\pref{sm11Dhole}, yielding the Matrix string-boosted
IIB hole transition curve at
\bb\label{hp2}
S\sim V^{-6}\ .
\ee

\sect{Perturbative 4+1d SYM phase} 
The scaling of
the equation of state is fixed by
dimensional analysis and yields
\bb\label{pertSYM4}
E\sim\lk(\frac{\r11}{N} \frac{1}{\lp^2}\re) V N^{1/2} S^{5/4}\ .
\ee
This phase borders that of the D4 branes and D0 branes.

The final phase diagram is that
of the $(2,0)$ on $T^4\times S^1$, or, as we see from the LC black hole
phase, that of LC M theory on $T^4$.

\subsection{The $(2,0)$ theory on $T_4/Z_2 \times S^1$}

Inspired by the previous discussion of the phase structure of the $(2,0)$
on $T^4\times S^1$, we further consider the phase structure of this theory
on $T^4/Z_2\times S^1$. This corresponds to a corner in the moduli space of 
$K3\times S^1$; particularly, in addition to considering
a square $T^4$, we will be ignoring phase dynamics associated
with the $16\times 4$ moduli that blow up the fixed points~\cite{ASPINWALLREV}.
Our parameter space is again two dimensional, entropy $S$ and the volume of
the $T^4$. There are only two novelties that arise,
both leaving the global structure of the phase
diagram unchanged, modifying only the interpretation of the various patches
of geometry. 

The first change arises from the effect of the orbifold on the duality 
transformations; we will obviously be driven into the other branch of
the web of dualities that converge onto M theory (\cf~\cite{POLCHV2}). 
We proceed from the
$\overline{11D}$ phase of the previous discussion, upward and counter-
clockwise on the phase diagram. We have M theory on a light-cone
circle times $T^4/Z_2$.
We reduce on $\r11$ to D0 branes in IIA living on the
$T^4/Z_2$ at~\pref{W11D0}. 
Under this orbifold, the massless spectrum has
positive parity eigenvalue. We T dualize on $T^4$ at~\pref{D0D4}, 
getting to the
patch of $D4$ branes in IIA wrapped
on $T^4/Z_2$. We remind the reader of the
transformation
\bb
T_{(4)}\beta_{(4)}T_{(4)}^{-1}=\beta_{(4)}\ ,
\ee
where we have used the properties of the reflection operator on the
spinors
\bb
\beta_i=\Gamma \Gamma^i\quad , \qquad
\beta_i^2=(-1)^{F_L}\quad , \qquad
\lk\{ \beta_i,\beta_j\re\}=0\ ,
\ee
with the T duality operation
reflecting the left moving spinors only. Here, $(-1)^{F_L}$
is the left moving fermion operator. 
We then M lift to M5 branes
in $\wtilde{M}$ theory on $T^4/Z_2\times S^1$ at~\pref{M5D4}. 
Next, we have to apply
the chain of dualities $M,T_{(3)},S$ near~\pref{mts}. 
From the M reduction we obtain
$\wtilde{D4}$ branes on $T^3/(-1)^{F_L}\Omega$. This is because the
M reduction along an orbifold direction yields the twist eigenvalues for
the massless spectrum
\bb
g_{\mu\nu}\ +;\ \ \phi\ +;\ \ B_{\mu\nu}\ -;\ \ C^{(1)}\ -;\ \ C^{(3)}\ +\ ,
\ee
while the world-sheet parity operator $\Omega$ acts on this spectrum as
\bb
g_{\mu \nu}\ +;\ \phi +;\ B_{\mu\nu}\ -;\ C^{(1),(2),(5),(6)}\ +;\ 
C^{(0),(3),(4),(7),(8)}\ -\ ,
\ee
and the action of $(-1)^{F_L}$ yields
\bb
\rm NSNS\ +\quad;\qquad RR -\ .
\ee
The T duality on $T^3$ brings us to $D1$ branes in IIB theory on
$S^1\times T^3/\Omega$, which is type I theory on $S^1\times T^3$.
This is because
\bb
T_{(3)} \beta_{(3)} (-1)^{F_L} \Omega T_{(3)}^{-1}=(-1)^{F_L}\Omega\ .
\ee
Finally, the S duality culminates in the geometry of $N$ black Heterotic 
strings smeared on the $T^3$. 
The Horowitz-Polchinski correspondence 
curve~\pref{corrloc} patches this phase onto that of the
Heterotic Matrix string phase, whose equation of state obeys
the scaling~\pref{matstringeos}. We thus verify the following previous
suggestions~\cite{BR,DIACGOMISMS,BANKSMOTL,MOTLSUSSKIND} 
from the perspective of Maldacena's conjecture:

\begin{itemize}
\item Heterotic Matrix string theory emerges in the UV of the $(2,0)$ theory.

\item Heterotic Matrix strings can be described via the $O(N)$ SYM of type I
D strings
\end{itemize}

\noindent
The structure of the phase diagram has not changed, but the labelling of
some of the phases has.
The additional symmetry structure of the orbifold background entered our
discussion trivially; the critical behaviors are unaffected.

To complete the discussion, we need to address a second change 
to the $T^4$ compactification.  The
localization transitions, say the one occurring at~\pref{D0loc}, 
are of a somewhat different nature than the ones encountered
earlier.  Localized black geometries on orbifold
backgrounds are unstable toward collapse toward the nearest fixed point;
by virtue of being above extremality, there are static forces, and by
virtue of the symmetry structure of the orbifold, there is no balance of
forces as in the toroidal case. 
It is then most probable that the localized D0 branes sit at the
orbifold points, with their black horizons surrounding the
singularity. The most natural geometry is the one corresponding to 16 black
D0 geometries distributed among the singularities, yielding a non-singular
geometry outside the horizons.

\subsection{\label{litstr}Little strings and fivebranes on $T^5$}

In this section, we study the thermodynamics of five branes wrapped on
a square $T^5$.
The notation is as before; we express all equations in terms of
the parameters of a LC M theory on $T^5$. The structure 
of the phase diagram for $V>1$ is similar to the one already encountered.
We will therefore not discuss the $D0$, $\overline{D0}$, $W11$,
$\overline{W11}$, $11D BH$, $\overline{11D} BH$, and perturbative 5+1d
phases except for noting that the only changes to our previous discussion
are to equations~\pref{W11D0}, \pref{sm11Dhole}
and \pref{pertSYM4}, which become respectively
\bbb
S&\sim& V^{-5/2}N^{3/2}\label{sm11Dhole2a}\\
E&\sim& \lk(\frac{\r11}{N}\frac{1}{\lp^2} \re) V^{5/2} S^{3/2}
\label{sm11Dhole2b}\\
E&\sim&\lk(\frac{\r11}{N} \frac{1}{\lp^2}\re) V N^{3/5} S^{6/5}
\label{sm11Dhole2c}\ .
\eee
We start from the $D5$ geometry and move counter-clockwise on the phase
diagram.

\sect{The M5 phase ($\wtilde{M5}$)} 
This phase consists of seven geometrical patches. 
For two of these, the $\overline{D0}$ and $\overline{W11}$,
we refer the reader to \cite{MSSYM123}. The relevant harmonic functions are
\bb
H=1+\frac{q^2}{r^2}\quad ,\qquad
h=1-\frac{r_0^2}{r^2}\ ,
\ee
with
\bb
q^2\sim \frac{\lp^4}{\r11^2}\frac{N}{V^5}\quad , \qquad
r_0^4\sim \frac{S^2}{N} \lp^4 V^{-5}\ .
\ee
The phase is described by the equation of state
\footnote{Note that we have kept track of the exact numerical coefficient
for this equation of state for later use.}
\bb
E=\frac{1}{2\pi} \frac{\r11}{\lp^2} S N^{-1/2} V^{5/2}\ ,
\label{Hagedorn}
\ee
characteristic of a string in a Hagedorn phase.
Our starting point is the {\it black D5 geometry ($D5$)} given by
\bbb\label{D5geom}
  ds_{10}^2&=&H^{-1/2}\lk( -h dt^2+dy_{(5)}^2\re) +
	H^{1/2} \lk( h^{-1} dr^2+ r^2 d\Omega_3^2\re)\nonumber\\
  e^\phi&=&H^{-1/2}\ .
\eee
The patch is parameterized by
\bb
\gs= \frac{\lp}{\r11} V^{-5}\quad , \qquad
\alp= \frac{\lp^3}{\r11}\quad , \qquad
y_{(5)}\approx \frac{\lp^2}{\r11} V^{-1}\ .
\ee
The relevant restrictions are:

\begin{itemize}

\item The Horowitz-Polchinski correspondence principle is satisfied for
\bb
S>V^{15/2} N^{-1/2}\ .
\ee
Beyond this point, we sew onto the perturbative 5+1d SYM phase whose
equation of state is given by~\pref{sm11Dhole2c}.

\item Requiring that the coupling at the horizon is small yields
\bb
S<V^{15/2} N^{3/2}\ .
\ee
We then S dualize to the geometry of black $NS5$ branes in the IIB theory.

\item The condition of large $T^5$ cycles at the horizon requires
\bb
S>V^{3/2} N^{3/2}\ .
\ee
Otherwise, we T dualize on the $T^5$ and obtain the geometry of smeared 
$D0$ branes $\overline{D0}$.
\end{itemize}

\noindent
{\it The black NS5 geometry ($NS5B$)} is the S dual of~\pref{D5geom}
\bbb\label{ns5b}
  ds_{10}^2&=&-h dt^2 + dy_5^2 
	+ H\lk(h^{-1} dr^2+r^2 d\Omega_3^2\re)\nonumber\\
  e^\phi&=&H^{1/2}\ ,
\eee
and the new asymptotic moduli are
\bb
\gs= \frac{\r11}{\lp} V^5\quad , \qquad
\alp= \frac{\lp^4}{\r11^2} V^{-5}\quad , \qquad
y_5 \approx \frac{\lp^2}{\r11} V^{-1}\ .
\ee
The relevant restrictions are:

\begin{itemize}

\item Requiring that the cycle size of the five $y$'s is greater than the
string scale yields the condition
\bb
V>1\ .
\ee
Beyond this point, we need to T dualize on the $T^5$ and we obtain the
geometry of black $NS5$ branes in a IIA theory.

\item The correspondence point is at
\bb
N\sim 1\ .
\ee
We note that the dual theory is the non-local
$(1,1)$ theory of IIB $NS5$ branes. At
low energy, it is described by 5+1d perturbative SYM.
\end{itemize}

\noindent
{\it The geometry of the black $NS5$ branes in IIA theory ($NS5A$)} is
the T dual of $NS5B$ (equation~\pref{ns5b})
\bbb
  ds_{10}^2&=&-h dt^2+dy_{(5)}^2+H \lk( h^{-1} dr^2
	+r^2 d\Omega_3^2\re)\nonumber\\ 
  e^\phi&=&H^{1/2}\ .
\eee
The parameters of the IIA theory are
\bb
\gs= \frac{\r11}{\lp} V^{-5/2}\quad , \qquad
\alp= \frac{\lp^4}{\r11^2} V^{-5}\quad , \qquad
y_{(5)} \approx \frac{\lp^2}{\r11} V^{-4}\ .
\ee
The new restrictions are:

\begin{itemize}

\item The correspondence point occurs for
\bb
N\sim 1\ .
\ee
The dual theory is the non-local
$(2,0)$ theory of IIA $NS5$ branes, related to the
$(1,1)$ theory we encountered above via a T duality on the $T^5$. 

\item Requiring small coupling at the horizon yields
\bb
S>V^{-15/2} N^{3/2}\ .
\ee
Otherwise, we lift to an $\wtilde{M}$ theory 
and obtain smeared $\wtilde{M5}$ branes.
\end{itemize}

\noindent
{\it The smeared black M5 geometry ($\wtilde{M5}$)} is 
described by the metric
\bb\label{M5geom}
ds_{11}^2=H^{2/3} \lk( d{\tilde{x}_{11}}^2+h^{-1} dr^2+r^2 d\Omega_3^2\re)
+H^{-1/3} \lk( dy_{(5)}^2-h dt^2\re)\ .
\ee
The parameters of the $\wtilde{M}$ theory are
\bb
\tr11= \lp V^{-5}\quad , \qquad
{\tlp}^3= \frac{\lp^5}{\r11^2} V^{-10}\quad , \qquad
y_{(5)}\approx \frac{\lp^2}{\r11} V^{-4}\ .
\label{Mtildeparams}
\ee
The new restrictions are:

\begin{itemize}

\item Requiring that the size of $\tilde{x}_{11}$ as measured at the horizon 
is smaller than the size of the object gives
\bb
S> V^{-15/2} N^{1/2}\ .
\ee
Otherwise, we localize \`{a} la Gregory-LaFlamme on $\tilde{x}_{11}$ 
to the geometry of localized $M5$ branes.

\item The correspondence condition yields
\bb
S> V^{-15/2} N^{-3/2}\ ,
\ee
which is rendered irrelevant by the previous condition.

\item Requiring that the cycle size of the $y$'s at the horizon
is bigger than the Planck scale yields
\bb
S> V^{3/2} N^{3/2}\ .
\ee
Beyond this point, we reduce on one of the cycles of $T^5$ to a IIA theory.
We find that we need to further T dualize on the remaining $T^4$. The
resulting geometry of black $D0$ branes is found strongly coupled at the
horizon; we therefore lift to another $\what{M}$ theory, and we
have the geometry of black $\what{M}$ waves smeared on the $T^4$.
\end{itemize}

\noindent
{\it The geometry of smeared waves in the $\what{M}$ theory 
($\what{M}\overline{W11}$)} is given by
\bbb
ds_{11}^2& = &\lk( H-1\re) \lk( d{\hat{x}_{11}}-dt\re)^2+
d{\hat{x}_{11}}^2-dt^2+H^{-1} \lk(1-h\re) dt^2
+dy_{(4)}^2\nonumber \\
& + &d{\tilde{x}_{11}}^2+h^{-1} dr^2+r^2d\Omega_3^2\ .
\eee
The parameters of the $\what{M}$ theory are
\bb
\hr11= \r11\quad , \qquad
\hlp= \lp V^{-2}\quad , \qquad
y_{(4)}\approx \lp V^{-2} \quad , \qquad
\tr11 \approx \lp V^{-5}\ .
\ee
The relevant restrictions are:

\begin{itemize}

\item Requiring that the cycle size of
$\tilde{x}_{11}$ at the horizon is bigger
than the Planck scale $\hlp$ yields
\bb
V<1\ .
\ee
Otherwise, we reduce along $\tilde{x}_{11}$ to a IIA theory, T dualize on
the $T^4$, and M lift back to the original LC M theory with Planck scale
$\lp$ and five torus moduli $V \lp$.

\item The system would localize on $\tilde{x}_{11}$ unless
\bb
S>V^{-15/2} N^{1/2}\ .
\ee
We then have localized waves in $\what{M}$ theory which are still smeared
on the remaining $T^4$.

\item We find that the cycle sizes of the four $y$'s as measured at the
horizon are of order the Planck scale $\hlp$. 

\item The system would localize on the $T^4$ unless
\bb
S> V^{-3/2} N^{1/2}\ .
\ee
This condition is never realized because of the other restrictions.

\item The system can localize on $\hat{x}_{11}$ unless
\bb
S>N\ .
\ee
Otherwise, we 
collapse to the geometry of an $11D$ black hole in LC $\what{M}$
theory; this black hole is still smeared on the $T^4$ and on
$\tilde{x}_{11}$.

\end{itemize}

\noindent
We thus conclude the discussion of this phase comprised of seven patches. 
We have two non-local theories sitting on top of the phase, 
the $(2,0)$ theory and the $(1,1)$ theory,
related by a T duality, and bounded by three curves due to
finite size effects, and one curve due to the correspondence principle.

\sect{The black M5 phase ($M5$)}
This phase consists of two patches.
{\it The M5 patch (M5)} is the localized version
of~\pref{M5geom}
\bb
ds_{11}^2=H^{2/3} \lk( h^{-1} dr^2+r^2 d\Omega_4^2\re)
+H^{-1/3} \lk( dy_{(5)}^2-h dt^2\re)\ ,
\ee
with the changes
\bb
q^3\sim \frac{\lp^5}{\r11^2} N V^{-10}\quad , \qquad
r_0^5\sim \lp^5 N^{-1} S^2 V^{-10} \ .
\ee
The equation of state becomes
\bb
E\sim \frac{\r11}{\lp^2} S^{6/5} V^4 N^{-3/5}\ ;
\label{bm5eos}
\ee
In other words, $S\sim N^{1/2}(y_{(5)} E)^{5/6}$
in the parameters \pref{Mtildeparams} of this patch;
this equation of state is
characteristic of a 5+1d gas, as one expects for the
theory on the M5-brane at large volume and sufficiently low energy.
The new restrictions are:

\begin{itemize}

\item The correspondence point is now at
\bb
N\sim 1\ .
\ee

\item Reduction on the $y$'s along the discussion for the smeared
$\wtilde{M5}$ branes encountered above occurs at
\bb
S\sim N^{4/3}\ .
\ee
We then emerge into the phase of $\what{M}W11$ black waves.
\end{itemize}

\noindent
{\it The geometry of $11D$ black waves ($\what{M}W11$)} is obtained
via localization on $\tilde{x}_{11}$ of the smeared geometry 
$\what{M}\overline{W11}$. The resulting
phase is still smeared on the $T^4$. It can however further localize at
\bb\label{sisn}
S\sim N
\ee
along $\hat{x}_{11}$ into a smeared $11D$ LC black hole $\what{11D} BH$.
The condition of localization on the $T^4$ is however
\bb\label{sis1}
S<N^{1/2}\ ,
\ee
and therefore never arises due to~\pref{sisn}.

\sect{The smeared $11D$ LC black hole phase ($\what{11D} BH$)}
This phase is described by the equation of state
\bb\label{hole4}
E\sim \frac{\r11}{\lp^2} N^{-1} V^4 S^{8/5}\ .
\ee
It is smeared on the $T^4$ but localized on $\tilde{x}_{11}$.
Minimizing its Gibbs energy as given by~\pref{hole4} with respect
to that of the hole smeared on $\tilde{x}_{11}$~\pref{sm11Dhole2b} yields
the transition curve
\bb
S\sim V^{-15}\ .
\ee
Localization on the $T^4$ occurs at $S\sim 1$, and therefore is not
seen on our phase diagrams. This can be seen by matching equation~\pref{hole4}
with
\bb
E\sim \frac{\r11}{\lp^2} N^{-1} V^4 S^{16/9}\ ,
\ee
\ie\ the equation of state of the totally localized hole in the
$\what{M}$ theory.

This completes the phase diagram obtained from the $D5$ system. 
The structure can be verified by using the various equations of state.
We conclude by noting that there are
several different interpretation of this diagram. It is that of the
$(2,0)$ theory; it is that of the $(1,1)$ related to the latter by
T duality; but it also encompasses the phase structure of LC M theory
on $T^5$. Various previous observations regarding Matrix theory 
on $T^5$ are thus confirmed~\cite{SEIBLITTLE,BRS} via the Maldacena conjecture.

\subsection{The $D6$ system}

\sect{The Taub-NUT Phase}
This phase consists of 8 patches. The harmonic functions are
\bb
H=1+\frac{q}{r}\quad , \qquad h=1-\frac{r_0}{r}\ ,
\ee
with
\bbb
r_0&\sim& S^{2/3} N^{-1/3} V^{-2} \lp\nonumber\\
q&\sim& \frac{\lp^3}{\r11^2} N V^{-6}\ .
\eee
The equation of state is
\bb
E\sim \frac{\r11}{\lp^2} S^{2/3} N^{-1/3} V^4\ .
\ee

Our starting point is {\it the black $D6$ geometry ($D6$)}, given by
\bbb
\label{fstrength}
  ds_{10}^2&=&H^{-1/2} \lk( -h dt^2+dy_{(6)}^2\re)
	+H^{1/2} \lk( h^{-1} dr^2+r^2d\Omega_2^2\re)\nonumber\\
  e^\phi&=&H^{-3/4}\nonumber\\
  F_{rty_1\cdots y_6}&=&\del_r H^{-1}\ .
\eee
The parameters of this IIA theory are
\bb
\alp= \frac{\lp^3}{\r11}\quad , \qquad g_s=
\lk( \frac{\lp}{\r11}\re)^{3/2} V^{-6}\quad , \qquad
y\approx \frac{\lp^2}{\r11} V^{-1}\ .
\ee

The various restrictions are:

\begin{itemize}
\item Weak coupling at the horizon requires
\bb
S<N^2 V^6\ .
\ee
Otherwise, we lift to a Taub-NUT geometry in eleven dimensions.

\item The correspondence point is at
\bb
S>V^6\ .
\ee
Perturbative 6+1d SYM emerges beyond this point.

\item T-duality on the $y_{(6)}$ must be applied unless
\bb
S>N^2\ .
\ee
Otherwise, we have the geometry of smeared $\overline{D0}$ branes.
\end{itemize}

\noindent
{\it The black Taub-NUT patch in $\overline{M}$ theory ($\overline{M}TN$)}
is given by the geometry
\bb
ds_{11}^2=H\lk( h^{-1} dr^2+r^2 d\Omega_2^2\re)
+H^{-1} \lk(dx_{11}-Ad\phi\re)^2-hdt^2+dy_{(6)}^2\ .
\ee
We have introduced a gauge potential $A=(1-\cos \theta)N/2$ 
locally ($\theta\neq 0$) for the magnetically
charged 2-form dual to~\pref{fstrength}. In the Maldacena limit,
this is an eleven dimensional 
ALE space with $A_{N-1}$ singularity.
The parameters of this $\overline{M}$ theory are
\bb
\tr11= \frac{\lp^3}{\r11^2} V^{-6}\quad , \qquad
\tlp^3= \frac{\lp^6}{\r11^3} V^{-6}\quad , \qquad
y_{(6)}\approx \frac{\lp^2}{\r11}V^{-1}\ .
\ee
The relevant restrictions are:
\begin{itemize}
\item The correspondence point:
\bb
S>N^{-1} V^6\ .
\ee
This is seen to be irrelevant.

\item The $T^6$ must be bigger than the Planck scale:
\bb
V>1\ .
\ee
Otherwise, we have to reduce along one of the cycles to a IIA
theory, and T dualize along the other five cycles to a IIB Taub-NUT
geometry. We then need to S dualize, and T dualize again on the five
torus; finally, we lift to a Taub-NUT geometry in an $\wtilde{M}$.
Instead of following this path, we will map the $V<1$ region from the
$\overline{D0}$ geometry.
\end{itemize}

\noindent
As mentioned above, we now pick up the trail from {it the $\overline{D0}$
patch}. This patch localizes on the $T^6$ to the $D0$ geometry for
\bb\label{locd6}
S> V^{9/2} N^{1/2}\ ,
\ee
and lifts to {\it an M theory wave $\overline{W11}$} for
\bb
S< V^{-6} N^2\ .
\ee
The latter localizes on the $T^6$ at~\pref{locd6}. For
\bb
V< 1
\ee
we need to reduce the $\overline{W11}$ geometry along one of the cycles
of the $T^6$, and T dualize on the other five. We find the coupling at the
horizon is bigger than one, so we S dualize, and find that the $T^5$ is
substringy. We T dualize again and find that the resulting IIA wave
geometry is strongly coupled at the horizon. We therefore, and finally,
lift to a black wave geometry in an $\wtilde{M}$ theory. The chain of
dualities is then $M,T_5,S,T_5,M$. The new {\it $\wtilde{M}$ wave geometry
($\wtilde{M}\overline{W11}$)} is parametrized by
\bb
z_{(6)}\approx \lp V^{-5}\quad , \qquad \tlp= \lp V^{-4}\quad , \qquad
x_{11}\approx \r11\ .
\ee
The $\wtilde{M}$ circle is one of the $z_{(6)}$, and the wave is along
$x_{11}$.
This geometry localizes on the $T^6$ for
\bb\label{locd6t}
S<V^{-9/2} N^{1/2}\ .
\ee
The $T^6$ at the horizon is bigger than $\tlp$ for $V<1$, and $x_{11}$
at the horizon is bigger than $\tlp$ for
\bb
S<V^6 N^2\ .
\ee
Otherwise, we reduce to a $\wtilde{IIA}$ 
theory along $x_{11}$ to the geometry of smeared $D0$ branes 
$\wtilde{\overline{D0}}$. The curvature at the horizon is small with
respect to the Planck scale for
\bb
S>V^{-3} N^{1/2}\ ,
\ee
which is rendered irrelevant by the other considerations.

\noindent
{\it The smeared $D0$ geometry $\wtilde{\overline{D0}}$} is parametrized
by
\bb
\gs= \lk(\frac{\r11}{\lp}\re)^{3/2} V^6\quad , \qquad
\alp= \frac{\lp^3}{\r11} V^{-12}\quad , \qquad
z_{(6)}\approx \lp V^{-5}\ .
\ee
A T duality on the $T^6$ takes us to the $\wtilde{D6}$ geometry for
\bb
S>N^2\ ,
\ee
and localization on the $T^6$ occurs for~\pref{locd6t}.

\noindent
{\it The $\wtilde{D6}$ geometry} is parametrized by
\bb
\gs= \lk(\frac{\lp}{\r11}\re)^{3/2}\quad , \qquad
\alp= \frac{\lp^3}{\r11} V^{-12}\quad , \qquad
z_{(6)}\approx \frac{\lp^2}{\r11} V^{-7}\ .
\ee
This has a correspondence point at
\bb
S\sim V^{-6}\ ,
\ee
and lifts to a Taub-NUT geometry in some $\what{M}$ theory for
\bb
S>V^{-6} N^2\ .
\ee

\noindent
{\it The Taub-NUT geometry $\what{M}TN$} obtained from the $\wtilde{D6}$
patch is parametrized by
\bb
\hr11= \frac{\lp^3}{\r11^2} V^{-6}\quad , \qquad
\hlp= \frac{\lp^2}{\r11} V^{-6}\quad , \qquad
z_{(6)}\approx \frac{\lp^2}{\r11} V^{-7}\ .
\ee
It patches onto the $\overline{M}TN$ geometry at $V\sim 1$ via a chain
of five dualities $M,T_5,S,T_5,M$ discussed above.

We note the symmetry of the diagram about $V\sim 1$. The remaining phases
were encountered in the previous SYM examples; there is a phase of
localized black D0 branes, a LC black hole, a smeared LC black hole, and
a 6+1d perturbative SYM phase. In the D6 system, each of these phases
has a mirror phase about the $V\sim 1$. The structure can be
further verified by matching the energies, at fixed entropy, of the
various phases. This completes the phase diagram for the $D6$ system, 
shown in Figure~\ref{SYM6fig}. We note that:
\begin{itemize}
\item The
gravitational coupling does not vanish in the $\overline{M}TN$ and
$\what{M}TN$ patches, whereas it does for all the other
patches of the diagram.

\item For $p=5,6$ diagrams involving p+1d SYM, the energy decreases
for higher entropies, unlike the $p<5$ cases; \ie\ the specific
heat is negative.

\end{itemize}

\subsection{\label{winding}Little strings with winding charge}

We will map here the thermodynamic phase diagram of $Q_5$ fivebranes
and $Q_1$ strings.
Our starting point is the $D1D5$ geometry.

\sect{Black fivebranes and strings}
This phase consists of 12 patches. We start with {\it the $D1D5$ geometry
($D1D5$)} given by
\bbb
  ds_{10}^2&=&\lk(H_1 H_5\re)^{-1/2} \lk( -h dt^2 + dx_5^2\re)
	+H_1^{1/2} H_5^{-1/2} dx_{(4)}^2
	+\lk(H_1 H_5 \re)^{1/2} \lk( h^{-1} dr^2
	+r^2 d\Omega_3^2\re)\nonumber\\
  e^\phi&=&H_1^{1/2} H_5^{-1/2}\nonumber\\
  F_{rtx_5}&=&\partial_r\lk(1+\frac{\rho_1^2}{r^2}\re)^{-1}\quad , \qquad
  F_{\theta_1\theta_2\phi}=2\rho_5^2 
  \lk(\varepsilon_3\re)_{\theta_1\theta_2\phi}
\label{D1D5metric}
\eee
The harmonic functions are given by
\bb
H_i=1+\frac{r_i^2}{r^2}\quad i=1,5\quad , \qquad
h=1-\frac{r_0^2}{r^2}\ ;
\ee
the charge radii of the branes $\rho_i$, $i=1,5$, are related
to the parameters $r_i$ by
\bb
\rho_1^2=\lk(2\pi\re)^4 \gs \alp^3 \frac{\lk(k q\re)^{1/2}}{V_4}\quad ,
\qquad
\rho_5^2=\gs k^{1/2} q^{-1/2} \alp\quad , \qquad
\rho_i^2=2r_i\sqrt{r_0^2+r_i^2}\ .
\label{rhor}
\ee
Here we make a distinction between the antisymmetric tensor
field strength's harmonic functions and
those of the metric, since we will be interested below in the numerical 
coefficients of some of the equations of state;
the extremal limit corresponds to $r_0\rightarrow 0$ with the
$\rho_i$ held fixed. 
For scaling purposes, we can write $\rho_i=r_i$ in the Maldacena limit. 
We also have traded the two integers
$Q_1$ and $Q_5$ for the new variables $k$ and $q$
\bb
Q_1\equiv \sqrt{k q}\quad , \qquad Q_5\equiv \sqrt{\frac{k}{q}}\ .
\ee
This IIB theory is parameterized by $\gs$, $\alp$, and lives on $T^4\times S^1$;
the $T^4$ is square with volume $V_4$, and we define\cite{MALDASTROMADS3}
\bb
v\equiv \frac{V_4}{\alp^2}\quad ,\qquad g_6\equiv \frac{g_s}{v^{1/2}}\ .
\ee
The $S^1$ is compact with radius $R_5$. The Maldacena limit corresponds
to
\bb
\alp\rightarrow 0 \mbox{ with }\frac{r}{\alp}\ ,\ R_5\ ,
\ g_6\mbox{ and }v\mbox{ held fixed.}
\label{D1D5maldalimit}
\ee
This reduces the geometry above to $AdS_3 \times S^3\times T^4$. The 
1+1d boundary theory is conformal with central charge $c=6 k$.
The gravitational coupling in our conventions is
\bb
G_{10}=\lk( 2\pi\re)^7 \gs^2 \alp^4\ .
\ee
From the area law, we have
\bb
S=\frac{\lk(2 \pi\re)^4}{G_{10}} r_1 r_5 r_0 R_5 V_4\ ,
\ee
or
\bb\label{r0s}
r_0\sim S \gs \alp k^{-1/2} v^{-1/2} R_5^{-1}\ .
\ee
The ADM mass is
\bb
M=\frac{\lk(2\pi\re)^3}{2} \frac{R_5 V_4}{G_{10}} \lk[ 3 r_0^2+
2\lk( r_1^2+r_5^2\re)\re]\ ,
\label{ADMenergy}
\ee
yielding the equation of state
\bb\label{Egeom}
E=\frac{S^2}{8\pi^2 k R_5} 
\ee
characteristic of a 1+1d conformal field theory~\cite{STROMVAFA}.

The various restrictions on the $D1D5$ geometry are:

\begin{itemize}

\item The Horowitz-Polchinski correspondence principle dictates
\bb
g_6>k^{-1/2}\ .
\ee
Beyond this point, the 1+1d conformal theory takes over.
Its equation of state is fixed by conformal symmetry; 
using Cardy's formula~\cite{CARDY,KUTSEIB} and the central 
charge $6 k$, we find precisely~\pref{Egeom}, as expected.

\item Requiring that the coupling at the horizon is small yields
\bb
g_6<q^{-1/2}\ .
\ee
Otherwise, we S dualize to the geometry of $NS5$ branes and fundamental
strings.

\item Requiring the the $T^4$ as measured at the horizon
is big with respect to the string scale gives
\bb\label{qg1}
q>1\ .
\ee
Otherwise, we apply a T duality on the $T^4$, and exchange the roles
of $Q_1$ and $Q_5$. Without loss of generality, we restrict our
attention to $q>1$ only. We also note that $q<k$; the upper bound
corresponds to $Q_5=1$. We therefore have
\bb
1<q<k\ .
\ee

\item Requiring that the size of $x_5$ as measured at the horizon 
is bigger than the string scale gives
\bb
S>g_6^{-1/2} k^{3/4}\ .
\ee
Otherwise, we T dualize to the geometry of smeared $D0D4$ branes.

\end{itemize}

\noindent
{\it The smeared $D0D4$ geometry ($\overline{D0D4}$)} is given by
\bbb\label{smd0d4}
  ds_{10}^2&=&-\lk(H_1 H_5\re)^{-1/2} f dt^2
	+ H_1^{1/2}H_5^{-1/2} dx_{(4)}^2
	+\lk(H_1 H_5\re)^{1/2} \lk( dx_5^2+f^{-1} dr^2 
	+ r^2 d\Omega_3^2\re)\nonumber\\
  e^\phi&=&H_1^{3/4} H_5^{-1/4}\ .
\eee
The parameters of this IIA theory become
\bb
\tgs= \gs \alp^{1/2} R_5^{-1}\quad , \qquad
\talp= \alp\quad , \qquad
x_{(4)}\approx \alp^{1/2} v^{1/4}\quad , \qquad
x_5 \approx \alp R_5^{-1}\ .
\ee
The restrictions are:

\begin{itemize}

\item Small curvature at the horizon yields the condition
\bb
g_6>k^{-1/2}\ .
\ee
This will be rendered irrelevant by the subsequent conditions.

\item Small coupling at the horizon requires
\bb\label{m5wlift}
S>g_6^{1/2} k^{3/4} q^{1/2}\ .
\ee
Otherwise, we lift to the geometry of smeared boosted $M5$ branes.

\item Requiring that the size of $x_5$ as measured at the horizon is
smaller than the transverse size of the object yields
\bb\label{l5}
S>g_6^{-1} k^{1/2}\ .
\ee
Beyond this point, the system localizes \`{a} la Gregory-LaFlamme along
$x_5$, and we have the geometry of localized $D0D4$ branes.

\item Finally, a large $T^4$ is associated with the condition~\pref{qg1}.

\end{itemize}

\noindent
{\it The smeared boosted M5 geometry ($\overline{M5W}$)} is the M lift of
the $\overline{D0D4}$ geometry (equation~\pref{smd0d4}) at~\pref{m5wlift}
\bbb
ds_{11}^2& = &H_1^{-1} H_5^{-1/3} \lk(-f dt^2+ H_1 dx_{(4)}^2\re)
+H_5^{2/3} \lk(dx_5^2+f^{-1} dr^2+r^2d\Omega_3^2\re)\nonumber \\
& &+ H_1 H_5^{-1/3} \lk( dx_{11}-\lk( H_1^{-1} -1\re) dt\re)^2\ .
\eee
The parameters of this M theory are
\bb
\r11= \gs \alp R_5^{-1}\quad , \qquad
\lp^3= \gs \alp^2 R_5^{-1}\quad , \qquad
x_{(4)}\approx \alp^{1/2} v^{1/4}\quad , \qquad
x_5 \approx \alp R_5^{-1}\ .
\label{M5Wparams}
\ee
The restrictions are:

\begin{itemize}

\item The correspondence principle requires
\bb
S>g_6^{-1} q^{1/2}\ .
\ee
This condition is rendered irrelevant by the others for $q<k$. At $q\sim k$,
it coincides with the localization condition on $x_5$ we will find shortly.

\item Requiring that the size of $x_5$ as measured at the horizon is 
bigger than the Planck scale yields
\bb
S< g_6^{-1} k^{3/4} q^{-1/4}\ .
\ee
Otherwise, we reduce along $x_5$ to a IIA theory and to the geometry of
boosted $NS5$ branes.

\item Requiring that the size of
the $T^4$ as measured at the horizon is bigger than the Planck
length gives
\bb\label{threedual}
S> g_6^{1/2} k^{3/4} q^{-1/4}\ .
\ee
Otherwise, we reduce to a IIA theory along one of the cycles of the $T^4$.
We find as always that the other three cycles are substringy and T dualize
along them. Finally, the resulting boosted $D1$ geometry is seen to be
strongly coupled at the horizon, and we S dualize to the geometry of 
boosted IIB fundamental strings smeared on $x_5$.

\item The localization condition on $x_5$ is as for the $\overline{D0D4}$ 
phase~\pref{l5}.

\end{itemize}

\noindent
{\it The geometry of $NS5$ branes and fundamental strings ($NS5FB$)}
is obtained from the $D1D5$ geometry via S duality
\bbb
ds_{10}^2&=&H_1^{-1} \lk(-f dt^2+H_1 dx_{(4)}^2\re) 
+H_1^{-1} dx_5^2+ H_5 \lk( f^{-1} dr^2+r^2 d\Omega_3^2\re)\nonumber\\
e^\phi&=&H_1^{-1/2} H_5^{1/2}\ .
\eee
The parameters of the IIB theory are
\bb
\tgs= \gs^{-1} \quad , \qquad
\talp= \gs \alp\quad , \qquad
x_{(4)}\approx \alp^{1/2} v^{1/4}\quad , \qquad
x_5 \approx R_5
\ee
The restrictions are:

\begin{itemize}

\item Small curvature at the horizon requires
\bb
k>q\ ,
\ee
which is trivially satisfied.

\item Large $x_5$ at the horizon requires
\bb
S>k^{3/4} q^{1/4}\ .
\ee
Otherwise, we T dualize along $x_5$ and emerge into the geometry
of boosted $NS5$ branes in IIA theory; the latter was encountered
from the $\overline{M5W}$ phase via an M reduction along $x_5$.

\item Large $T^4$ at the horizon requires
\bb
g_6<1\ .
\ee
Otherwise, we T dualize along the $T^4$, yielding to a similar system
with altered asymptotic moduli.

\end{itemize}

\noindent
{\it The boosted black IIB string geometry ($\overline{F1WB}$)} is
obtained from the $\overline{M5W}$ patch by a chain of three dualities as
described after equation~\pref{threedual}; this gives the metric
\bbb
ds_{10}^2&=&-\lk(H_1 H_5\re)^{-1} f dt^2+ H_1 H_5^{-1} d{\hat{x}}_{11}^2
+ dx_{(3)}^2+f^{-1} dr^2 +r^2 d\Omega_3^2+dx_5^2\nonumber\\
e^\phi &=& H_5 ^{-1/2}\ .
\eee
The parameters of the IIB theory are
\bbb
& {\hat g}_\str &= \gs^{-1} \alp^{-1/2} v^{3/4} R_5 \quad , \qquad
{\hat\alpha}'= \gs^2 \alp^2 v^{-1} R_5^{-2}\nonumber \\
& x_{(3)} &\approx \gs \alp v^{-1/2} R_5^{-1}\quad , \qquad
x_5 \approx \alp R_5^{-1}\quad , \qquad
\hat{x}_{11}\approx \gs \alp R_5^{-1}\ .
\eee
The restrictions are:

\begin{itemize}

\item Small curvature at the horizon requires
\bb\label{sk12}
S>k^{1/2}\ .
\ee
This will be rendered irrelevant by other restrictions.

\item Large $x_5$ as measured at the horizon requires
\bb
g_6<1\ .
\ee
Otherwise, we T dualize along $x_5$, and obtain a similar
geometry.

\item Localization on $x_5$ occurs unless condition~\pref{l5} is satisfied.

\item Localization on $x_{(3)}$ occurs unless condition~\pref{sk12}
is satisfied. This is irrelevant in view of~\pref{l5}.

\item Requiring that $\hat{x}_{11}$ as measured at the horizon
is bigger than the string scale yields equation~\pref{qg1}.

\item Small coupling at the horizon requires the reverse of~\pref{threedual}.

\item And finally, we note that the geometry is at the self-dual point
for the three cycles $x_{(3)}$.

\end{itemize}

\noindent
{\it The geometry of boosted $NS5$ branes of the IIA theory ($NS5WA$)}
is obtained from the $NS5FB$ patch via T duality or the $\overline{M5W}$
via M reduction. The only relevant restriction is that of large $T^4$
at the horizon. This occurs for
\bb\label{smallg6}
g_6<1\ .
\ee
Otherwise, we have the T dual, and identical, geometry
with different asymptotic moduli.

We have completed the boosted $M5$ phase up to the condition~\pref{smallg6}.
We note that all duality transformations along $g_6\sim 1$ leave the
geometries unchanged, and change the asymptotic moduli. It is easy then
to check that venturing into domains with $g_6>1$ yields a mirrored
picture of the phase diagram about $g_6\sim 1$. Our six patches have 
six other mirror geometries across the $g_6\sim 1$ line.
We therefore see a signature of a strong-weak symmetry 
$g_6\rightarrow 1/g_6$ in the dual theory, which
is T-duality of the little string. As we scan through $1<q<k$,
at the lower bound the phase structure is such that, via dualities
exchanging $Q_1$ and $Q_5$, a mirrored phase diagram for $q<1$ emerges; for
the upper bound, the geometrical vacua across the diagram break down 
via the correspondence principle. These comments carry over to the
other phases, which we describe next.

\sect{The black localized boosted $M5$ phase ($M5W$)}
The localization transition along $x_5$ yields the change in the harmonic
functions
\bb
f\rightarrow 1-\frac{r_0^3}{r^3}\quad , \qquad
H_i\rightarrow 1-\frac{r_i^3}{r^3}\ ,
\ee
with
\bbb
r_1^3&\sim& \frac{\alp^2 \gs}{v R_5} k^{1/2} q^{1/2}\nonumber\\
r_5^3&\sim& \frac{\alp^2 \gs}{R_5} k^{1/2} q^{-1/2}\ .
\eee
The expression \pref{r0s} for the entropy
is unaffected by this transition, unlike all
other cases encountered here and in~\cite{MSSYM123}.
The equation of state of the localized phase becomes
\bb\label{D0D4eos}
E\sim \frac{g_6}{R_5} \lk(\frac{S}{k^{1/2}}\re)^3\ .
\ee

There are three patches. {\it The localized boosted fundamental string
($F1WB$)} is obtained from the $\overline{F1WB}$ patch by localization
on $x_5$. The relevant restrictions are:

\begin{itemize}

\item Small curvature at the horizon requires
\bb\label{SK12}
S>k^{1/2}\ .
\ee
At this point, we emerge in a Matrix string phase carrying two charges.
More on this later.

\item Localization on $x_{(3)}$ occurs unless~\pref{SK12} is satisfied.
This is similar to what we saw in the 4+1d SYM case.

\item Large $x_{11}$ at the horizon requires~\pref{qg1}.

\item Small coupling at the horizon necessitates
\bb
S< k^{2/3} q^{-1/6}\ .
\ee
Beyond this point, we apply the chain $S, T_{(3)}, M$ to patch onto
the localized boosted $M5$ geometry. The reverse of this chain
was described in the smeared case above.

\item Finally, the geometry is at the self-dual point for the $x_{(3)}$
cycles.

\end{itemize}

\noindent
{\it The localized boosted black $M5$ brane geometry ($M5W$)} 
has the same parameters as \pref{M5Wparams}.
It is subject to one additional non-trivial condition, 
that of M reduction along $x_{11}$ unless
\bb
S<k^{2/3} q^{1/3}\ .
\ee
We then emerge into the geometry of localized $D0D4$ branes.

\noindent
{\it The localized $D0D4$ brane geometry ($D0D4$)} is subject
to the following additional condition; its curvature at the
horizon is small when
\bb
S<k\ .
\ee
Otherwise, the dual geometrical description breaks down. Comparing the
equations of state~\pref{D0D4eos} and \pref{Egeom}, we find that we do
not have a match at $S\sim k$. This is identical to the situation encountered
in all the SYM cases at $S\sim N^2$. There is a non-trivial transition
at this point through a phase with zero specific heat. On the 
1+1d gas side, $S\sim k$ is where the thermal wavelength becomes the size of
the box $R_5$; the dynamics is then frozen into a quantum mechanics. 

\sect{The BPS Matrix String}
At $S\sim k^{1/2}$, the emerging phase is that of the Ramond
ground states of the conformal theory, 
which are those of a BPS Matrix string. 
The situation can be compared to the matrix string transition
of the $M5$-brane on $T^4\times S^1$ discussed in Section \ref{6dqft}.
There, we found a correspondence curve at
$S\sim V^{-3}Q_5^{1/2}$, beyond which a perturbative
string description should be valid.  
At the transition, the ratio of the cycle sizes at the horizon of
the $T^4$ and the $S^1$ was
$(y_4/\tr11)^2\sim V^6\sim Q_5/S^2$; in particular,
since $V\ll 1$, the dynamics is effectively one-dimensional.
In the localized ($M5W$) phase of the D1-D5 system, we have 
$(y_4/\r11)^2\sim H_1^{-1}(\frac{\alp v^{1/2} R_5^2}{g_s^2\alp^2})
\sim Q_1^{-1}$ at the transition $S\sim k^{1/2}$,
which is again of order $Q_5/S^2$.
We conclude that the two transitions are the same.
In the present case, the emerging phase is BPS; a perturbative string
carrying both winding $Q_5$ and momentum $Q_1$ quanta 
obeys the Virasoro constraints
\bbb
  E^2&=&(Q_1\ls/R)^2+(Q_5 R/2\ls)^2 + 2N_L+2N_R\nonumber\\
  k&=&Q_1Q_5=N_L-N_R\ ;
\eee
when \eg\ the left oscillator level $N_L<<N_R$, there are of order
$k^{1/2}$ states, and the system becomes BPS-saturated at $N_R=0$.

\sect{\label{DLCQsect} Comments on DLCQ of the M5-brane}
As mentioned in the introductory summary, the limit $Q_5$ fixed,
$Q_1\gg Q_5$, is relevant to the DLCQ description of the 
M5-brane~\cite{ABKSS,SETHISEIBERG,SETHI,GANORSETHI}.
%
%
In terms of the D1-D5 parameters, the DLCQ parameters are
\bbb
  \frac{\lp^2}{\r11}
	&=&\ls\left(\frac{R_5}{\ls\gs}\right)^{\frac13}\nonumber\\
  \frac{x_{(4)}}{\lp}
	&=&v^{1/4}\left(\frac{R_5}{\ls\gs}\right)^{\frac13}
	\equiv \nu_4^{1/4}\nonumber\\
  \frac{x_5}{\lp}
	&=&\frac{\ls}{R_5}
	\left(\frac{R_5}{\ls\gs}\right)^{\frac13}
	\equiv L\ .
\label{DLCQparams}
\eee
Converting \pref{Egeom}
to DLCQ parameters \pref{DLCQparams}, we find
\bb
  S=2\pi(Q_5/L)^{\frac12}\lp M\ ,
\ee
which is indeed the equation of state of a Hagedorn string with
tension proportional to $L/Q_5$, as has been
seen previously from several related points of view
~\cite{MALDAFIVE,AHARONYBANKS}.
It is a nontrivial check that this equation 
of state agrees precisely with Equation \pref{Hagedorn}
when we use the parameters \pref{Mtildeparams} 
of the $\wtilde{M5}$ phase
\footnote{The light-cone scaling was determined in \cite{AHARONYBANKS};
our contribution is a check that the precise numerical
coefficient agrees.}.

The same exercise can be repeated for the localized ($M5W$) phase.
The equation of state \pref{D0D4eos} in DLCQ parameters,
again assuming light-cone kinematics $E_{LC}\sim  M^2/2P$,
takes the form
\bb
  S\sim Q_5^{1/2} Q_1^{1/6} (\nu_4^{1/4}\,\lp M)^{2/3}\ .
\label{DLCQloc}
\ee
In terms of scaling, this equation of state
is the energy-entropy relation
of a 2+1d gas (extensive in the box size $\nu_4^{1/4}$), 
although it is difficult to explain the dependence
on $Q_1$ and $Q_5$.  
A natural candidate for the object being observed
here is an excited $M2$-brane embedded in the $M5$-brane
(which is indeed one of the bound states of M-theory).
The $Q_1$ dependence appears to violate Lorentz invariance;
it would be interesting to understand why light-cone
kinematics does not work in the low-energy, low-entropy regime;
and why the low-entropy phase is not a boosted version
of the 5+1d gas found for the $M5$-brane
in Equation \pref{bm5eos}.

\subsection{\label{specflowsec}Spectral flow and rotating black holes}

We now turn to a discussion of angular momentum in the D1-D5 system.
As pointed out in Section \pref{specflowintro},
spectral flow is an adiabatic twisting of boundary conditions in
the full string theory before
the Maldacena limit; the near-horizon limit maps the
twist onto the spectral flow operation in the superconformal algebra.
On the geometry side, a point on the unitarity diagram 
(Figure \ref{unitarity}) in the NS sector, far
from the boundary and at high conformal weight, is described by the
BTZ black hole geometry (independent of the fermion boundary conditions)
~\cite{BTZ,MALDASTROMADS3,MARTINECADSMM,MARTINECGEOM}, in a space which is
asymptotically locally $AdS^3\times S^3\times \MM_4$; 
in the R sector, such a point represents the near horizon geometry
of a rotating D1D5 system~\cite{CVETICLARSEN}
(due to the shift in conventions between canonical 
definitions of NS and R sector quantum numbers).
The isometry $SO(4)=SU(2)_L\times SU(2)_R$ of the transverse
$S^3$, combines with the $(4,4)$
supersymmetry generators and the $SL(2,R)\times SL(2,R)$
symmetry of $AdS_3$, to yield two copies of the $\NN=4$ superconformal 
algebra; a gauge transformation in
$SU(2)$ can be used to shift the boundary conditions on the supercharges,
yielding an isomorphism between the NS and R sectors~\cite{SCHWIMSEIB}.
The charges under the Cartan subgroups 
of each of the two $SU(2)$'s parametrize the
angular momenta of the rotating D1D5
or BTZ hole geometries. 
The subalgebra of concern is then two copies 
of the $\NN=2$ superconformal algebra 
with two $U(1)$ R-symmetry generators 
$J^3_{L,R}=\hf q^{U(1)}_{L,R}\equiv j$ that implement
the spectral flow. 
We restrict our attention to equal
left and right $U(1)$ charges.

Consider first the NS sector.
The qualitative features of the density of states about $j=0$
were discussed in~\cite{BDHM}.
There are several phases.  Consider the regime of
sufficiently large effective coupling $\geff^2\equiv g_6^2 k>1$; 
in the present notation, for $E R_{AdS}=2h\greaterapprox k$ 
one is in the BTZ black hole phase~\cite{BDHM} with
$S\sim(kh)^{1/2}$.\footnote{The standard conventions for the
BTZ metric, where length and time scales are referred
to the $AdS$ curvature radius, differ from those of the D1D5
geometry encountered in the R sector,
where scales are often referred to the scale $R_5$.
Matching the asymptotics of the metrics
yields the relation $E_{NS} R_{AdS}\sim E_{R} R_5\sim h$,
where $R_{AdS}^4\sim G_6 k$. We write subsequent equations in terms of the
invariant conformal weight $h$ to avoid confusion.}
For $k \geff^{-3/2}\lessapprox h\lessapprox k$,
there is a phase of 5+1d Schwarzschild black holes
because the horizon localizes on $S^3$;
the entropy is of order $S\sim k^{-1/3}h^{4/3}$.
The lower bound is the correspondence point; thus,
for $\geff^{3/2}\lessapprox h\lessapprox k\geff^{-3/2}$,
there is a Hagedorn phase, with $S\sim h\geff^{-1/2}$.  
Finally, for $h\lessapprox \geff^{3/2}$,
the system is in a supergravity gas phase, with $S\sim h^{5/6}$.
At weak coupling $\geff<1$, the 5+1d Schwarzschild phase
and the supergravity gas phase disappear.

Consider next the R sector (\ie\ $j\approx k/2$), 
and define $h'\equiv h-\coeff14 k$ and $j'=j-\coeff12 k$
as the energy and angular momentum in R sector
conventions. We first focus on the regime
$\geff>1$, \ie\ the middle part of the diagram in Figure~\ref{D1D5fig}. For
$\geff^{-2} k \lessapprox h'$, we have the black $D1D5$ system. 
For $0 < h'\lessapprox \geff^{-2} k $, we have the M5W or D0D4
phase localized on $x_5$. 
Finally, at $h'\sim 0$, the BPS Matrix string cuts the diagram
at finite and large entropy $S\sim k^{1/2}$. 
For $\geff<1$, we have an additional
phase with entropy $S \sim k$ for $\geff k \lessapprox h' \lessapprox k$
squeezed between the D1D5 and D0D4 phases. 
On the phase diagram of Figure~\ref{D1D5fig}, this
corresponded to the horizontal line segment at $S\sim k$.
As argued in~\cite{PEETROSS}, we see that
the D1D5 system without angular momentum does not localize
on the $S^3$ at low energies, whereas the stationary BTZ hole in the NS
sector does undergo such a localization~\cite{BDHM}.
The spectral flow map adiabatically relates
the states of these two sectors; the differing phase structures
obtained at zero charge in
the NS and R sectors (the latter flowing to \eg\ $j=k/2$
in the NS sector) then implies that the
spinning D1D5 system must undergo a localization transition
on the $S^3$ at a critical value of the angular momentum.  
We next analyze the possibility for such a transition.

The equation of state of the rotating D1D5 phase 
can be extracted from the corresponding geometry~\cite{CVETICLARSEN}, and
is given by~\pref{denstates}
\bb
  S_{\rm BTZ}^2\sim kh'-j^{\prime\,2}\ .
\ee
This phase should callapse at a critical value of $j'$ 
to a 5+1d black hole localized and spinning on the $S^3$. 
Angular momentum is introduced in this phase by spinning up the black
hole along an orbit on the equator of $S^3$ with momentum
$p\sim j'/R_{AdS}$; kinematic relations
and the Schwarzschild equation of state then imply
\bb
  S_{\rm 6d}\sim k^{-1/3} \lk( {h'}^2-j^{\prime\,2}\re)^{2/3}
	\rightarrow k^{-1/3} {h'}^{4/3}\ ,
\label{S6d}
\ee
where, in the last step, we have taken the non-relativistic limit $h'\gg j'$;
we will see that this is justified.
In the relativistic limit the hole approaches extremality;
one obtains a gravitational wave on $S^3$ with $h'\sim j'$,
thus matching onto the BPS spectrum of supergravity states.
This regime occurs near the boundary of the unitarity plot, 
where the Hagedorn or gas phase takes over the 5+1d black hole.
The localization on the $S^3$ will then occur if $S_{\rm BTZ}<S_{\rm 6d}$
at a given energy, \ie\ 
\bb
\frac{h'}{k} < \lk(\frac{j'}{k}\re)^2 +\lk(\frac{h'}{k}\re)^{8/3}\ .
\ee
For $h'< k$, \ie\ for the horizon size smaller than the size of the $S^3$,
we can ignore the last term, and we have the condition
\bb\label{locj}
h'< \frac{j^{\prime\,2}}{k}\ .
\ee
Note that for $j'$ near zero, 
the corresponding localization condition cannot
be met~\cite{PEETROSS}.
For $j'\sim k$, however, this equation can be satisfied:
The direct analysis in the NS sector $j'=\hf k$ 
shows that a localized phase exists at large enough $\geff$.
We conclude that the $D1D5$ system indeed localizes on the $S^3$ at
a critical value of the angular momentum.  Note that
our uncertainty in the location of the transition
is due to the fact that
it is sensitive to the numerical accuracy of the equations, 
not just the scalings of the thermodynamic parameters 
(and therefore lies beyond the scope of our geometrical analysis).
Continuing to lower conformal weights in the R sector,
the equation of state of the rotating D0D4 phase is given by
\cite{CVETICYOUM}
\bb
S^2\sim k\lk(\frac{h'}{g_6}\re)^{2/3} - j'^2\ .
\ee
In the NS sector, the rotating BTZ hole localizes on $S^3$
as the energy is lowered; the equation of state is roughly \pref{S6d}
(without the primes).
As extremality is approached, the spinning black hole reaches the
correspondence point and becomes a large fundamental string
carrying angular momentum
\bb
  S_{\rm Hag}\sim [\geff^{-1/2} h - j^2]\ .
\ee
Below this, there should be a transition where a supergravity
gas extremizes the free energy. 
The phase structure about the NS sector $j\sim 0$ 
should sew onto the phase structure about the R sector $j\sim k/2$
in some intermediate regime.
Most of the above formulae are not invariant under spectral flow; 
they are determined
in an analysis about zero angular momentum
relative to the NS or R sectors, and may be corrected
by large gravitational back-reaction when $j\sim k$.
We leave an analysis of these effects to future work.  

For $\geff<1$, the picture is slightly different 
(see Figure \ref{hversusj}).
There is no 5+1d black hole or supergravity
gas phase in the NS sector.
The phase labeled SQM has an entropy which is
energy independent $S\sim k$. As mentioned above, it correponds to the 
horizontal line at $S\sim k$ in Figure~\ref{D1D5fig}. We again defer a
detailed analysis to future work. 

It is a curious fact that, for finite $k$, the spectral
flow relation between the NS and R sectors implies
an IR cutoff in the spectrum of particle states in
the latter.  In the NS sector, the eigenmodes of the
free scalar wave equation have a natural gap in the spectrum
of order $1/R_{AdS}\sim (g_6^2 k\alp)^{-1/2}$; however,
in the R sector, the free spectrum is continuous.
Nevertheless, in the full quantum theory, spectral
flow from one sector to the other implies that the
finite density of states in the NS sector gives a finite
R sector density of states; in other words, finite $k$ generates
an effective IR cutoff.  This cutoff disappears in
the classical $k\rightarrow\infty$ limit, as one
sees for example in the fact that the number of BPS states
in the R sector is $O(\sqrt{k})$.
This feature is a property of all Maldacena-inspired definitions
of quantum gravity (\ie\ using finite $N$ dual nongravitational systems)
where the dual theory is in finite volume (\eg\ a torus); the
finite density of states due to the IR cutoff in the gauge theory
imposes an effective cutoff at large radius in the supergravity --
even though the classical wave equations in such geometries
can have continuous spectra.  It would be interesting to understand
this phenomenon better (it is not obviously related to the UV-IR
correspondence of \cite{SUSSWIT}).

\newpage
{{\Large \bf Acknowledgments:}}
We wish to thank 
J. Harvey,
D. Kutasov,
F. Larsen,
and
A. Lawrence
for helpful discussions.
This work was supported by DOE grant DE-FG02-90ER-40560.

\providecommand{\href}[2]{#2}\begingroup\raggedright\endgroup

\end{document}